\documentclass[aps,prl,twocolumn,superscriptaddress,longbibliography]{revtex4-2}
\usepackage{amsmath,amssymb}
\usepackage{graphicx,color}
\usepackage{times}
\usepackage[pdftex,colorlinks,breaklinks]{hyperref}
\hypersetup{linkcolor=blue,citecolor=blue,filecolor=dullmagenta,urlcolor=blue, 
 pdftitle={Parametric correlations in Dissipative Quantum Chaos},
 pdfauthor={Fyodorov, Lacroix-A-Chez-Toine}, pdflang={English} }

\renewcommand{\Im}{\textrm{Im}}
\renewcommand{\Re}{\textrm{Re}}

\newcommand{\Tr}{\mathrm{Tr}}

\newcommand{\beginsupplement}{%
        \setcounter{table}{0}
        \renewcommand{\thetable}{S\arabic{table}}%
        \setcounter{figure}{0}
        \renewcommand{\thefigure}{S\arabic{figure}}%
        \setcounter{equation}{0}
        \renewcommand{\theequation}{S\arabic{equation}}%
        \setcounter{section}{0}
        \renewcommand{\thesection}{S\arabic{section}}%
        \setcounter{page}{1}
}

\begin{document}
\title{Parametric correlations in non-Hermitian quantum chaos: random matrix approach}
\author{Yan V. Fyodorov and Bertrand Lacroix-A-Chez-Toine}
 \affiliation{Department of Mathematics, King’s College London, London, WC2R 2LS, United Kingdom}
\date{\today}
\begin{abstract}
Motivated by the surge of interest  in statistics of non-Hermitian random matrices as a framework for description of universal characteristics of dissipative chaotic quantum many-body systems,  we address the problem of characterizing the parametric correlations of spectral densities.  Considering parameter-dependent ensemble of complex Ginibre matrices we derive an explicit, closed-form expression for the parametric number covariance in the systems of symmetry class $\mathbf{A}$ for eigenvalues in a circular domain containing on average a finite number of eigenvalues in the spectral bulk.  This behavior is expected to be universal, as further supported by numerical evidence for the real Ginibre ensemble, non-Hermitian Bernoulli Wigner matrices and bi-unitarily invariant ensembles. We also discuss a relation between parametric correlations of spectral densities and the distribution of the so-called eigenvector non-orthogonality factor, which attracted considerable interest in recent years. 
\end{abstract}
\makeatletter\let\MyTitle\@title\makeatother 
\maketitle

Bohigas-Giannoni-Schmit (BGS) \cite{BGS1984} conjecture asserts that the theory of large Hermitian random matrices 
(RMT) represents the most natural reference framework for identifying, characterizing, and quantifying universal phenomena in isolated quantum systems dominated by effects of chaos and disorder. When the Hamiltonian of such a system depends on a tunable parameter, e.g. the Aharonov-Bohm flux of magnetic field piercing the sample or gate potential, one may be interested in understanding the universal features of response of the system spectral characteristics, in particular,  energy levels and eigenfunctions, to a change in the parameter value \cite{gaspard1990parametric,goldberg1991parametric,akkerm_montamb_92,sierant2019fidelity}. 
 If the parameter change is {\it mesoscopic}, i.e. strong enough to move the energy levels considerably, but still weak enough so that nontrivial energy correlations are not completely washed out, the use of perturbative diagrammatics allows to recover the universal response features successfully \cite{szafer1993universal}, and can be reproduced using asymptotic analysis of Dyson Brownian motion \cite{beenakker1993brownian,frahm1995brownian} as well as by semiclassic techniques \cite{berrykeating94param}. On the other hand, to describe  microscopic regime of parameter change when the induced level motion is comparable to the mean level spacing requires instead employing the supersymmetric Efetov's nonlinear sigma-model approach \cite{simons1993universal1,simons1993universal2,simons_lee_altsh} or exact solutions of Sutherland-type Hamiltonians associated with random matrix Brownian motion models \cite{beenakker1994param, nagao1998multilevel}. 
The resulting theoretical expressions were successfully tested in several experiments \cite{exper_simons93,exper_bertelsen99,Experim99kudrolli,exper_savytskyy2001,
experDietz2020}. It was further realized that when the physical perturbation in the system's parameters is spatially confined, which can be modelled at RMT level by considering finite-rank perturbations,  universal response characteristics are modified \cite{aleiner1998shifts,marchetti2003universality,smolyarenko03param}. In recent years studies of parametric level motion have been extended from single-particle to many-body disordered systems, in particular as an indicator of many-body localization \cite{monthus2017mbl,maksymov2019energy,ChalkerParam_MBL21}.

In the last decade non-Hermitian effects arising in quantum many-body or optical wave systems due to inherent mechanisms of dissipation, loss or gain have attracted vigorous research interest, giving rise to an entirely new field, 
see Ref.~\cite{ashida2020non} for a review. In this context the BGS conjecture was extended already long ago by Grobe, Haake and Sommers (GHS) in \cite{grobe1988quantum} by suggesting that the theory of non-Hermitian random matrices should play the same reference role for non-conservative quantum chaotic systems as it played for isolated conservative counterparts. 
In recent years the utility of such an approach has been further supported by extensive studies of dissipative systems displaying many-body quantum chaos and eventually many-body localization phenomena \cite{akemann2019universal,denisov2019universal,jaiswal2019universality,can2019random,hamazaki2019non,sa2020spectrala,sa2020spectralb,sa2020complex,wang2020hierarchy,li2021spectral,tarnowski2021random,lange2021random,garcia2022symmetry,kulkarni2022lindbladian,ghosh2022spectral,suthar2022non,cipolloni2023entanglement,sa2023symmetry,de2022non,de2023non,ghosh2023eigenvector,orgad2024dynamical,jisha2024universality,richter2025integrability,almeida2025universality,rufo2025quantum,wold2025experimental,chirame2025open}.  Despite the fact that precise conditions of applicability and universality of GHS conjecture remain 
a matter of controversy and active discussion \cite{villasenor2024breakdown,villasenor2025correspondence,ferrari2025dissipative,mondal2026transient},
it very essentially invigorated the interest in spectral characteristics of non-Hermitian matrices.

In particular, numerical evidence combined with heuristic arguments \cite{hamazaki2020universality} suggest that there are only three types of local spectral correlations for complex eigenvalues in the spectral bulk. 
The simplest representatives for these three  universality classes are complex Ginibre (class ${\bf A}$), complex symmetric (class AI$^\dag$), and complex self-dual (class AII$^{\dag}$) Gaussian random matrices. While Ginibre matrices have been thoroughly investigated analytically for decades \cite{byun2025progress}, the systematic mathematical analysis of the two other non-standard classes has been started only very recently, see \cite{akemann2025complex,kulkarni25,forrester2025dualities,akemann2025spectral,kurilov2025density}. 

Motivated by all these recent developments it seems timely to start building the theory of parametric correlations in complex spectra of non-conservative quantum chaotic systems, using the non-Hermitian RMT as the framework. 
To the best of our knowledge only the parametric correlations at mesoscopic scale has been considered in the mathematical non-Hermitian RMT literature \cite{bourgade2024fluctuations}, while our primary goal is to address microscopic scaling.  
To this end, in this paper we consider fluctuations of the simplest informative object, the number $n_{\alpha}(D)$ of complex eigenvalues in a domain $D\subset \mathbb{C}$ of the complex plane for a fixed value of the control parameter $\alpha$. Denoting the exact (empirical) eigenvalue density 
in the complex plane as $\rho_\alpha(z)$, we have by definition $n_{\alpha}(D)=N\int_{D}\rho_\alpha(z)d^2z$, where $N$ is the total number of eigenvalues. Our main goal is to analyze the variations of $n_{\alpha}(D)$ induced by a change in the control parameter $\alpha$. We consider $\alpha=1$ to be the reference (unperturbed) ensemble and assume that changes in the parameter leave the average density invariant $\left\langle\rho_{\alpha}(z)\right\rangle=\left\langle\rho_{1}(z)\right\rangle$. Here and henceforth the angular brackets stand for the disorder averaging. The simplest nontrivial characteristic is thus the  {\it parametric number covariance} defined as 
\begin{equation} \label{parnumbvar} 
{\rm Var}_n(\alpha,D)=\left\langle\left[n_{1}(D)-n_{\alpha}(D)\right]^2\right\rangle\,,
\end{equation}
analogues of which have been studied in the Hermitian case in \cite{goldberg1991parametric,vinayak2016param}. To evaluate such a covariance we actually first address the underlying parametric two-point (connected) correlation function of the spectral densities, 
namely
\begin{equation} \label{parcorden_def}
C_{\alpha}(z_1,z_2)=\left\langle \rho_{1}(z_1) \rho_{\alpha}(z_2)\right\rangle-\left\langle \rho_{1}(z_1)\right\rangle \left\langle\rho_{\alpha}(z_2)\right\rangle\,.
\end{equation}
Our main result is an explicit expression for the above object evaluated in the limit $N\to \infty$ for the parametric version of the complex Ginibre ensemble defined via 
\begin{equation}\label{pramGincompl}
X_{\alpha}=\alpha X_1+\sqrt{1-\alpha^2}X_2, \quad 0\le \alpha\le 1,    
\end{equation}
where $X_1$ and $X_2$ are two independent copies of the standard complex $N\times N$ Ginibre matrices with probability distribution function (PDF) $P(X)\propto \exp(-N{\rm Tr}(X X^{\dagger}))$. This normalization ensures that their eigenvalues in the limit $N\to \infty$ uniformly fill in the unit disk in the complex plane. The parametric dependence in Eq. \eqref{pramGincompl} leaves the PDF invariant and denoting $\alpha=e^{-t/2}$ it can be naturally related to a Brownian evolution in time $t$ of the complex Ginibre ensemble \cite{bourgade2024fluctuations}. The statistical properties of the ensemble are stationary in $\phi={\rm arccos}(\alpha)$. 
Anticipating that  as $N\to \infty$ in the chosen normalization the eigenvalues of $X_{\alpha_1}$ and $X_{\alpha_2}$ are non-trivially correlated on the microscopic scale as long as the difference $\alpha_1-\alpha_2$ remains of the order $N^{-1}$, we further use stationarity and concentrate on the vicinity of $\alpha=1$ introducing the scaled control parameter $v^2=2N(1-\alpha)<\infty$. The parametric correlation function of the two densities at points $z_1$ and $z_2$ situated inside the unit circle
with mutual distance of the order of mean level spacing: $|z_1-z_2|\sim N^{-1/2}$ and the center of mass $z_0=(z_1+z_2)/2$ 
can be expressed in terms of the limiting mean eigenvalue density at the center of mass $\left\langle\rho(z_0)\right\rangle=\pi^{-1}$ and the parameter $a= v^2 \left\langle {\cal O}(z_0)\right\rangle$, with $\left\langle {\cal O}(z_0)\right\rangle=1-|z_0|^2$. Namely, we find that the density correlation function is given by \cite{SM}
\begin{equation} \label{parcorden_lim1}
\lim_{N\to \infty} C_{\alpha=1-\frac{v^2}{2N}}(z_1,z_2)=\left\langle\rho(z_0)\right\rangle^2\,c_a(u), 
\end{equation}
where $u=\pi \left\langle\rho(z_0)\right\rangle N|z_1-z_2|^2$ and
\begin{align} \label{parcorden_lim2}
c_a(u)=&\frac{2a^2}{(a+u)^3}-e^{-(a+u)}\\ \nonumber
&-\left[a(2+a)\frac{\partial^2}{\partial a^2}+a^2\frac{\partial^3}{\partial a^3}\right]\frac{1-e^{-(a+u)}}{a+u}\;.
\end{align}
Note that despite having  the explicit values for $\left\langle\rho(z_0)\right\rangle$ and  $\left\langle{\cal O}(z_0)\right\rangle$ in the complex Ginibre ensemble, we find it convenient to retain more general notations for later use. 

Using that the Dirac delta distribution in the complex plane can be recovered as $\lim_{a\to 0}\frac{2a^2}{(a+\pi\nu|z|^2)^3}=\nu^{-1}
\delta^{2}\left(z\right)$, we see that taking $a\to 0$ in  \eqref{parcorden_lim1} faithfully reproduces the standard microscopic-scale correlation function of the limiting spectral densities in the complex Ginibre ensemble \cite{byun2025progress}.
As a somewhat less trivial limit induced by the same term one may notice that in the limit $v\to 0$
\begin{align}\label{limit_nonorthB}
\lim_{v\to 0} v^2 c_{v^2 \tilde a}\left(v^2\tilde u\right) =\frac{2{\tilde a}^2}{\left(\tilde a+\tilde u\right)^3}.   
\end{align}

In the End Matter section\ref{EM} we show that  \eqref{limit_nonorthB} with $\tilde u=\pi\left\langle\rho(z_0)\right\rangle|p|^2$ for a fixed $p\in \mathbb{C}$ and $\tilde a=\left\langle{\cal O}(z_0)\right\rangle$ contains valuable information about the so-called "diagonal non-orthogonality factor" describing a nontrivial relation between  left and right eigenvectors of non-Hermitian random matrices \cite{ChalkerMehlig1998}. Namely, let $\mathbf{l}_n$ and $\mathbf{r}_n$ be left and right eigenvectors of $X_1$ corresponding to the same eigenvalue $z_n(1)$, i.e. satisfying $X_1\mathbf{r}_n=z_n(1)\mathbf{r}_n$ and $\mathbf{l}_n^{\dag}X_1=z_n(1)\mathbf{l}^{\dag}_n$. It is standard to assume the left and right eigenvectors satisfy the bi-orthogonality condition: $\mathbf{l}^{\dag}_a\mathbf{r}_b=\delta_{ab}$. The  diagonal non-orthogonality factor is then defined as $O_{nn}=\left(\mathbf{l}_n^{\dagger}\mathbf{l}_n\right)\,\left(\mathbf{r}_n^{\dagger}\mathbf{r}_n\right)$. This quantity is one of the most useful indicators of matrix non-normality. Its simplest statistical characteristic is the so-called Chalker-Mehlig mean value defined as 
 \begin{equation}\label{CMmean}
 {\cal O}_1(z)=\left\langle \frac{1}{N^2}\sum_{n=1}^NO_{nn}\delta^{(2)}(z-z_n(1))\right\rangle_{X_1}
 \end{equation}
 with a well-defined limit as $N\to \infty$, which in the Ginibre case equals 
to   $\lim_{N\to \infty}{\cal O}_1(z)=\frac{1}{\pi}(1-|z|^2)$\cite{ChalkerMehlig1998}. In the End Matter section\ref{EM} we demonstrate how to use \eqref{limit_nonorthB} to recover the whole limiting distribution of  $O_{nn}$, which has been obtained before by very different methods  \cite{bourgadedoubach2020,fyodorov2018CMP}.  
In particular, we will see that the parameter $\left\langle{\cal O}(z_0)\right\rangle$
entering the parametric density correlation function must be simply related  to the Chalker-Mehlig mean value at the center of mass: $\left\langle{\cal O}(z_0)\right\rangle=\pi {\cal O}_1(z_0)$.
 
 A relation between the parametric spectral density correlation function and the diagonal non-orthogonality factor could be anticipated on general grounds, as the latter is responsible for enhanced linear (first order) sensitivity of eigenvalues of non-normal matrices, see  \cite{SM} for more detail. In fact relations similar in spirit have been used in Hermitian setting for addressing the distribution of energy level derivatives with respect to control parameters \cite{kravtsov1992kramers,fyodorov1994velocities,curvature_fyosom1995,fyodorov1995level}, the first derivatives known as "level velocities" (or currents) and second derivatives being "level curvatures". 
 It is worth however mentioning  a significant difference between quantum chaotic non-Hermitian and Hermitian systems. Namely, in the Hermitian case the first derivatives are Gaussian-distributed unless effects of eigenfunction localization become operative \cite{fyodmirl1995meso}, and only second derivatives are nontrivially (heavy-tail) distributed \cite{gaspard1990parametric,zakrzewski1993parametric,curvature_VOpp1994,curvature_VOpp1995}.
 At the same time in the non-Hermitian setting already the first order (linear in perturbation) eigenvalue shifts are heavy-tail distributed due to the effects of eigenfunction nonorthogonality \cite{fyodorovSavin2012,bourgadedoubach2020,fyodorov2018CMP}. 

Finally, it makes sense to consider the limit $u\gg 1$ of $c_a(u)$ which is natural to call "mesoscopic". 
Neglecting exponentially small terms $O(e^{-u})$ in \eqref{parcorden_lim2} and applying  straightforward differentiation one can rewrite the parametric density correlation function in the form

\begin{eqnarray} \label{lim_meso}
\lim_{N\to \infty}\frac{C_{\alpha=1-\frac{v^2}{2N}}(z_1,z_2)}{\left\langle \rho(z_0)\right\rangle^2}\approx\frac{6a^2}{(a+u)^4}-\frac{4a}{(a+u)^3}\nonumber  \\
=\frac{1}{\left(N\pi \left\langle \rho(z_0)\right\rangle\right)^2}\frac{\partial^2}{\partial z_1\partial \overline{z}_1}\frac{\partial^2}{\partial z_2\partial \overline{z}_2}\ln\left(a+u\right)\,,
\end{eqnarray}
where we remind that $u=N\pi \left\langle \rho(z_0)\right\rangle|z_1-z_2|^2$, which exactly matches the known expression for the kernel operator governing covariances in parametric linear eigenvalue statistics through the mesoscopic regime of wide eigenvalue separation: $1\ll u\ll N$ \cite{bourgade2024fluctuations}. In particular, it illustrates the deep connections between parametric eigenvalue fluctuations and the highly active area of logarithmically correlated random fields, see 
\cite{bourgade2024fluctuations} for a detailed exposition.

Having at our disposal the expressions \eqref{parcorden_lim1}-\eqref{parcorden_lim2} we are now in position 
to evaluate our main object of interest, the parametric number covariance \eqref{parnumbvar}. Setting $\alpha=1-\frac{v^2}{2N}$ for a fixed $N$, we present our $N\to \infty$ results for the simplest choice of the observation domain  which we choose as the circle of center $z_0$ in the interior of the unit circle ($|z_0|<1$) and radius $R/\sqrt{N}$.  For such an observation domain $D={\cal D}_{z_0}(R)=\{z\in\mathbb{C}\,,\,\,|z-z_0|\leq R/\sqrt{N}\}$ we will denote the number $n({\cal D}_{z_0}(R))$ of eigenvalues of $X_{\alpha}$ inside it as ${\cal N}_{z_0}(R)$.  For a finite $0<R<\infty$ the number ${\cal N}_{z_0}(R)$ is finite on average and reads $\left\langle {\cal N}_{z_0}(R)\right\rangle=R^2$ . With this choice 
the parametric number covariance is given in terms of two parameters: $a=v^2(1-|z_0|^2)$ and $y=R^2$ by \cite{SM}
\begin{equation}
\lim_{N\to\infty}{\rm Var}_n\left(1-\frac{v^2}{2N},{\cal D}_{z_0}(R)\right)=2y\left[\sqrt{\frac{a}{a+4y}}+{\cal V}(a,y)\right],   \label{scaling-limit} 
\end{equation}
where
\begin{align}
&{\cal V}(a,y)=\left(1-e^{-a}\right)e^{-2y}\left[I_0(2y)+I_1(2y)\right]   \label{parnummvar_result}  \\
&-a\int_0^{1} e^{-\tau\left(2y+a\right)}\left(2+a(1-\tau)\right)\tau\left[I_0(2y\tau)+I_1(2y\tau)\right]\,d\tau.\nonumber
\end{align}
Note that in the limit $a\to \infty$ with fixed $y<\infty$ correlations between the eigenvalues of matrices at two different values of parameter vanish, and the above expression reduces to the ratio of the variance to the average number of eigenvalues for a fixed value of the parameter given by ${\cal V}(\infty,y)+1=e^{-2y}\left[I_0(2y)+I_1(2y)\right]$, see Eq.(3.16) in \cite{byun2025progress}, and \cite{shirai2006large,lacroix2019rotating,fenzl2022precise,akemann2023universality} for derivations in several different but equivalent forms.

{\it Comparison with numerics and universality conjectures} 

\begin{figure}
    \centering
    \includegraphics[width=0.98\linewidth]{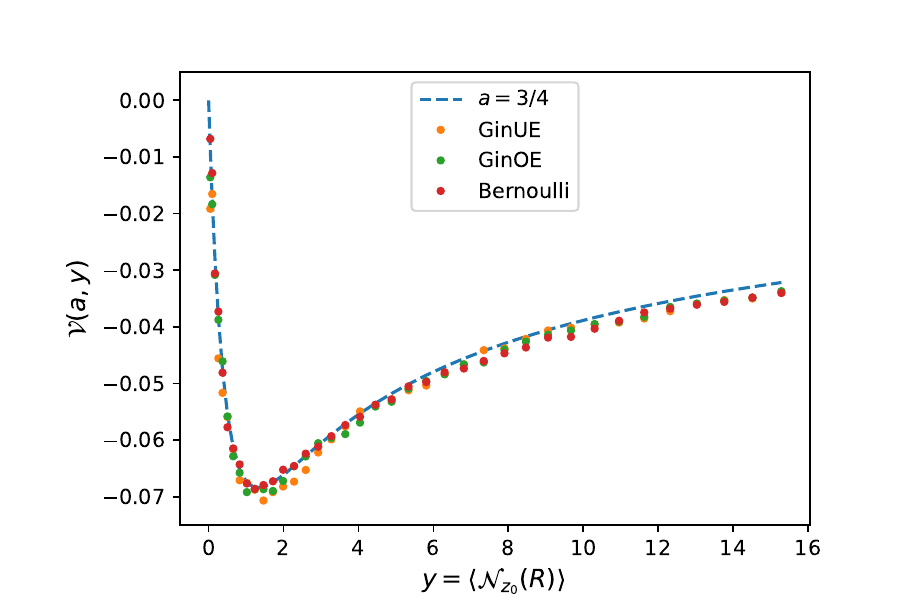}
    \caption{The rescaled parametric number covariance scaling function ${\cal V}(a,y)$ in the disk ${\cal D}_{z_0}(R)$ of centre $z_0=(1+i\sqrt{3})/4$ and varying radius $R$ is plotted as a function of the average number $y=\langle {\cal N}_{z_0}(R)\rangle=R^2$ for several random matrix ensembles. The good matching between the numerical data for the complex Ginibre ensemble GinUE (orange dots) and the theoretical curve (blue dashed line) provides evidence for the validity of our prediction. The matching between the theoretical prediction and numerical data from the real Ginibre ensemble GinOE (green dots) and from non-Hermitian Wigner matrix with Bernoulli distribution (red dots) is as good as for the complex Ginibre ensemble, providing evidence for the wider universality of our results. The simulations are done for a matrix size $N=400$ and $N_{\rm it}=10^5$ realizations.}
    \label{fig:IID}
\end{figure}

As a consequence of the well-known spectral universality of models in class A in the large size limit \cite{maltsev2025bulk}, the ratio ${\rm Var}({\cal N}_{z_0}(R))/\left\langle {\cal N}_{z_0}(R)\right\rangle$ between the variance and average number of eigenvalues in the disk ${\cal D}_{z_0}(R)=\{z\in\mathbb{C}\,,\,\,|z-z_0|\leq R/\sqrt{N}\}$ is universal and described by the scaling function ${\cal V}(\infty,y)+1$ with $y=\left\langle {\cal N}_{z_0}(R)\right\rangle$. We expect that the whole scaling function ${\cal V}(a,y)$ bears a similar universality for finite values of $a$ and appears in other models in this class. To state our conjecture on universality, we introduce an ensemble of disordered Hamiltonian $X$ of class A, drawn from a given PDF $P(X)$. The parametric dependence is then introduced via a global diffusive dynamics on the entries of the Hamiltonian, i.e. with all entries evolving with the same timescale. In this way, the Hamiltonian $X(t)$ becomes time dependent,  and we consider a stationary dynamics, i.e. preserving the marginal fixed-time PDF $P(X,t)=\left\langle \prod_{i,j=1}^N \delta(X-X_{ij}(t))\right\rangle=P(X)$. Note that the angular brackets throughout this section refer to an average over both the disorder and the random time-evolution. The time dependence of the entries naturally induces a dynamics on the eigenvalues $z_n(t)$ of the Hamiltonian and we denote $\rho(z,t)$ their empirical density at time $t$. The dynamics being stationary, fixed-time statistics are invariant and in particular  $\left\langle\rho(z,t)\right\rangle=\left\langle\rho(z)\right\rangle$. We similarly denote $n(D,t)=N\int_{D} \rho(z,t)d^2 z$ the number of eigenvalues in the domain $D$ at time $t$ and introduce its parametric covariance
\begin{equation}
    {\rm Var}_n(D,t)=\left\langle\left[n(D,0)-n(D,t)\right]^2\right\rangle\,.
\end{equation}
Our conjecture states that for models of class A and for a global, diffusive, stationary, isotropic dynamics in the short-time and large size limit with fixed $\tau=Nt$, the covariance of eigenvalues in the disk ${\cal D}_{z_0}(R)$ takes the universal scaling form
\begin{equation}
    \lim_{N\to \infty}{\rm Var}_n\left({\cal D}_{z_0}(R),\frac{\tau}{N}\right)=2y\left[\sqrt{\frac{a}{a+4y}}+{\cal V}(a,y)\right],\label{conjecture}
\end{equation}
where the parameters read $y=\left\langle {\cal N}_{z_0}(R)\right\rangle$ and
\begin{equation}
 a=\lim_{N\to \infty}\pi \left\langle\sum_{n=1}^N\left|z_n\left(\frac{\tau}{N}\right)-z_n(0)\right|^2\delta^2(z_n(0)-z_0)\right\rangle\;.\label{parameters} 
\end{equation}
Note that the identity $dz_n={\bf l}_n^{\dagger}dX {\bf r}_n$ allows to relate the parameter $a$ above with the Chalker-Mehlig diagonal correlator $O_1(z_0)$ (see \cite{SM}).
For concreteness, we give below explicit examples of dynamics and use them to test numerically our conjecture. First, for models with Gaussian disorder, the parametric dependence introduced in Eq. \eqref{pramGincompl} extends straightforwardly by interpolating all random couplings simultaneously with $\alpha=e^{-t/2}$. Next, we consider non-Hermitian Wigner matrices with i.i.d. complex entries $X_{ij}$ such that $P(X)=\prod_{i,j}p_N(\Re(X_{ij}))p_N(\Im(X_{ij}))$, with zero average, finite second $N\left\langle|X_{ij}|^2\right\rangle=2N\int x^2 p_N(x)dx =\sigma^2$ and finite higher moments. The dynamics considered consists in randomly updating the entries with a new realization, i.e.
\begin{equation}
    X_{ij}(t+dt)=\begin{cases}
     &X_{ij}(t)\;{\rm with\;probability}\;1-\gamma dt\\
     &\tilde X_{ij}\;{\rm with\;probability}\;\gamma dt
    \end{cases}\;,
\end{equation}
with $\tilde X_{ij}$ independent to $X_{ij}(t)$ with the same distribution. For these non-Hermitian Wigner ensembles, the expressions for the average of ${\cal N}_{z_0}(R)$ and the Chalker-Mehlig diagonal correlator are universal in the large $N$ limit \cite{osman2026universality} and coincide with the Ginibre results, i.e. $\left\langle {\cal N}_{z_0}(R)\right\rangle=R^2$ and $\pi O_1(z_0)=1-|z_0|^2$. Thus, the parameters read $y=R^2$ and $a=2\gamma \sigma^2\tau(1-|z_0|^2)$. Our numerical results for Bernoulli distribution of entries in Fig. \ref{fig:IID} show as good an agreement with the scaling form ${\cal V}(a,y)$ as in the Ginibre complex ensemble, providing evidence for the conjecture in Eq. \eqref{conjecture}.

Finally, we consider bi-unitarily invariant ensembles of matrices such that $P(X)\propto \exp(-N{\rm Tr}W(XX^{\dagger}))$. The Langevin dynamics of the form
\begin{equation}
    dX(t)=-\frac{1}{2}W'(X(t)X^{\dagger}(t))X(t)dt+\frac{dB(t)}{\sqrt{2N}}\label{Langdyn}
\end{equation}
where $dB(t)$ is a complex standard Brownian process on all entries of the matrix ensures that $P(X)$ is a stationary measure (see \cite{SM} for more details). In the short-time $t=\tau/N$ limit considered here, the first (drift) term on the right-hand side of Eq. \eqref{Langdyn} can be ignored and for all purposes $dX(t)\approx dB(t)/\sqrt{2N}$, which corresponds to a perturbation by the complex Ginibre ensemble. For that dynamics, the parameters $y=\pi \langle \rho(z_0)\rangle R^2$ and $a=\tau\pi O_1(z_0)$. For bi-unitarily invariant ensembles, the values of $\left\langle {\cal N}_{z_0}(R)\right\rangle=\pi \langle \rho(z_0)\rangle R^2$ and $O_1(z_0)=x(|z_0|)(1-x(|z_0|))/(\pi |z_0|^2)$ where $x(|z_0|)= \int_{|z|\leq |z_0|}\langle\rho(z)\rangle d^2z$ \cite{belinschi2017squared} can be computed in terms of the average density of eigenvalues $\langle\rho(z)\rangle=\pi^{-1}\left.\partial_u(u W'(u))\right|_{u=|z|^2}$. We perform additional numerical simulations to provide evidence of the validity of the conjecture for other bi-unitarily invariant ensembles (see \cite{SM} for details).

{\it Method of derivation}. Here we sketch the main steps of the derivation leading to  the expressions \eqref{parcorden_lim1}-\eqref{parcorden_lim2} in case of the complex Ginibre ensemble. Our approach extends to parametric correlations the method of the papers \cite{kanzieper2002replica,Kanzieper05}, based on
exploiting in a judicious way the exact integrability of replica theory related to determinants of Ginibre matrices.  
The evaluation starts with defining for a positive integer $n$ and the parameter value $\alpha$ the appropriate replica generating function 
\begin{equation}\label{genfundef}
{\cal Z}_n(z_1,\overline{z_1},z_2, \overline{z_2})=\left\langle \left|\det{z_1-X_{1}}\right|^{2n} \left|\det{z_2-X_{\alpha}}\right|^{2n}\right \rangle  
\end{equation}
where averaging is performed over two independent copies of Ginibre matrices $X_1$ and $X_2$.
Representing determinants in terms of integrals over anticommuting variables, the average can be efficiently 
performed by standard manipulations, see e.g. \cite{nishigaki2002replica}, with full details presented in \cite{SM}. 
As a result, one is able to represent the generating function \eqref{genfundef} via the following matrix integral over the set of non-Hermitian matrices $Q=\left(\begin{array}{cc} q_{11} &  q_{12}\\ q_{21} & q_{22}\end{array}\right)$, where each block $q_{ab}$ is a complex matrix of size $n\times n$:  
\begin{equation}\label{genfunint}
{\cal Z}_n(z_1,\overline{z_1},z_2, \overline{z_2})=\int e^{-N\mbox{\small Tr}QQ^{\dagger}}
\left[\det{\left(\begin{array}{cc} Q_{\alpha} & -i Z\\ -iZ^{\dagger} & Q_{\alpha}^{\dagger}
\end{array}\right)}\right]^N
\end{equation}
where we denoted $Q_{\alpha}=Q+(\sqrt{\alpha}-1)\left(\begin{array}{cc}0 & q_{12}\\q_{21}&0\end{array}\right)$
and $Z=\left(\begin{array}{cc} z_1{\bf 1}_n & 0\\0 & z_2{\bf 1}_n\end{array}\right)$.
In the large-N limit we employ the parameter scaling $\alpha=1-\frac{v^2}{2N}$ and after further introducing $z_1=z+\frac{\omega}{2\sqrt{N}},\, z_2=z-\frac{\omega}{2\sqrt{N}}$ one can extract the leading-order contribution to the integral in \eqref{genfunint} via a somewhat lengthy saddle-point evaluation presented in detail in \cite{SM}.  The resulting expression takes the form of a nonlinear sigma-model integral over $2n\times 2n$ unitary matrices $U_0\in U(2n)/U(n)\otimes U(n)$: 
\begin{align}\label{genfunint_sigma}
{\cal Z}_n(z_1,\overline{z_1},z_2, \overline{z_2})\propto& e^{-2n(1-|z|^2)\left(N+\frac{v^2}{4}\right)}\\
&\times\int d\mu(U_0) e^{\frac{|\omega|^2+v^2(1-|z|^2)}{4}\Tr\left[\Lambda U_0^{\dagger}\Lambda U_0\right]}\;,\nonumber
\end{align}
where we denoted $\Lambda=\left(\begin{array}{cc} {\bf 1}_n & 0\\0 & -{\bf 1}_n\end{array}\right)$.
As has been discovered by Kanzieper \cite{kanzieper2002replica,Kanzieper05}, for any integer $n$ this type of integral satisfies a positive, semi-infinite Toda Lattice Hierarchy and due to exact integrability can be mapped onto a solution of the fifth Painlevé transcendent equation, see further results in this direction in \cite{deano_simm2022}.
In such a solution the integer $n$ enters as an arbitrary real parameter, which allows to continue analytically 
and write the solution for small $n\ll 1$ in the form 
\begin{eqnarray}\label{genfunint_Painleve}
{\cal Z}_n(z_1,\overline{z_1},z_2, \overline{z_2})= e^{-2nN(1-|z|^2)}\Psi_n(x)
\end{eqnarray}
where we denoted $x=|\omega|^2+v^2(1-|z|^2)$ and 
\[
\Psi_n(x)=e^{n\frac{x}{2}-n^2\left[\ln{x}+\delta E(x)\right]+o(n^2)},
\]
with $\delta E(x)=\int_1^{\infty}e^{-xt}\left(\frac{1}{t}-\frac{1}{t^2}\right)\,dt$.
Having the above at our disposal, the parametric density correlation function \eqref{parcorden_lim1} is recovered in the replica limit $n\to 0$ by a straightforward but lengthy differentiation: 
\begin{equation}\nonumber
\left\langle \rho_{1}(z_1)\rho_{\alpha}(z_2)\right\rangle=\lim_{n\to 0}\frac{1}{\pi n^2}
\frac{\partial}{\partial z_2} \frac{\partial}{\partial z_1} \frac{\partial}{\partial \overline{z}_2} \frac{\partial}{\partial \overline{z}_1}   {\cal Z}_n(z_1,\overline{z_1},z_2, \overline{z_2})
\end{equation}

{\it Discussion and open problems}. 
The main challenge remains to develop more comprehensive understanding of parametric correlations in other symmetry classes. 
Experimentally it is also relevant to take into account locality/finite rank of the perturbation, which is expected to modify 
the results significantly. In fact such a difference between local vs. global perturbations have been indeed observed experimentally \cite{gros14} at the level of first-order parametric shifts of complex S-matrix poles \cite{fyodorovSavin2012} in billiards with wave-chaotic scattering. Other relevant directions may involve clarification of various aspects of parametric sensitivity in many body localized phase of quantum systems, extending the insights obtained for conservative many body systems in \cite{monthus2017mbl,maksymov2019energy,ChalkerParam_MBL21} to their dissipative counterparts. To this end it is worth noting that statistics of eigenvalues and eigenfunction non-orthogonality  has been numerically shown to be a sensitive indicator of eigenfunction localization in non-Hermitian  systems \cite{ghosh2022spectral,ghosh2023eigenvector}.

\begin{acknowledgments}
 Paul Bourgade and Giorgio Cipolloni are gratefully acknowledged for elucidating the content of the paper \cite{bourgade2024fluctuations}. The research of YVF has been supported by EPSRC grant UKRI1015 "Non-Hermitian random matrices: theory and applications". 
\end{acknowledgments}
\bibliography{parametric} 

\section{End Matter}\label{EM}
\setcounter{secnumdepth}{2}
\setcounter{equation}{0}
\renewcommand{\theequation}{A\arabic{equation}}
\setcounter{figure}{0}
\renewcommand{\thefigure}{A\arabic{figure}}
\subsection*{Appendix: Relating  \eqref{limit_nonorthB} to distribution of diagonal non-orthogonality factor}
According to the definition \eqref{pramGincompl} for small values of $v\ll 1$ we have 
$X_{\alpha=1-\frac{v^2}{2N}}=X_1+\frac{v}{\sqrt{N}}X_2+O(v^2/N)$.
This allows to use the first-order perturbation theory in $v\ll1$ to compute the leading-order shift in the eigenvalues $z_n(\alpha), n=1,\ldots,N$ of $X_{\alpha}$ as
\begin{equation}\label{eigshift}
z_n\left(\alpha=1-\frac{v^2}{2N}\right)-z_n(\alpha=1)=\frac{v}{\sqrt{N}}\frac{\mathbf{l}_n^{\dag}X_2\mathbf{r}_n}{\mathbf{l}^{\dag}_n\mathbf{r}_n}
\end{equation}
where $\mathbf{l}_n$ and $\mathbf{r}_n$ are left and right eigenvectors of $X_1$ corresponding to the eigenvalue $z_n(1)$. Now recalling the definition of the eigenvalue density for $X_{\alpha}$ as $\rho_\alpha(z)=\frac{1}{N}\sum_{n=1}^N\delta^{(2)}\left(z-z_n(\alpha)\right)$ we can represent the expression in \eqref{limit_nonorthB}
as 
\begin{equation}\label{e1_l} 
\lim_{v\to 0} v^2 C_{\alpha=1-\frac{v^2}{2N}}\left(z_1=z,z_2=z+p\frac{v}{\sqrt{N}}\right)
\end{equation}
\begin{equation}\label{e1_r}
=\frac{1}{N}\left\langle \sum_{n,m} \delta^{(2)}\left(z-z_n(1)\right) F_{nm}(p)\right\rangle_{X_1}
\end{equation}
where we defined
\begin{equation}\label{Fab}F_{nm}(p)=\lim_{v\to 0}\frac{v^2}{N}\left\langle \delta^{(2)}\left(\frac{pv}{\sqrt{N}}-
\left[z_m\left(1-\frac{v^2}{2N}\right)-z_n(1)\right]\right)\right\rangle_{X_2}.
\end{equation}
Due to generic non-degeneracy of eigenvalues of random matrices, the limit above is zero unless $n=m$. 
In the latter case  exploiting \eqref{eigshift} and evaluating the limit we find
\begin{equation}\label{e2}
F_{nm}(p)=\delta_{nm}\left\langle\delta^{(2)}\left(p-\mathbf{l}_n^{\dag}X_2\mathbf{r}_n\right)\right\rangle_{X_2}
\end{equation}
Since entries of $X_2$ are complex Gaussian-distributed variables independent of $X_1$, for a fixed realization of $X_1$ the quantity  $\mathbf{l}_n^{\dag}X_2\mathbf{r}_n$ is a mean-zero Gaussian-distributed complex variable. Hence for evaluating \eqref{e2} it is enough to know its second moment. A simple computation then gives 
\begin{equation}\label{e2a}
F_{nm}(p)=\delta_{nm}\frac{N}{\pi O_{nn}}e^{-\frac{N}{O_{nn}}|p|^2}, 
\end{equation}
where $O_{nn}=\left(\mathbf{l}_n^{\dagger}\mathbf{l}_n\right)\,
\left(\mathbf{r}_n^{\dagger}\mathbf{r}_n\right)$ is exactly the 
diagonal non-orthogonality factor introduced by Chalker and Mehlig \cite{ChalkerMehlig1998}. Further we introduce the joint probability density (JPD) of the eigenvalues and the corresponding non-orthogonal factors for the matrix $X_1$ as 
\begin{equation}\label{e3}
{\cal P}_N(z,\omega)=
\frac{1}{N}\left\langle\sum_{n=1}^N \delta^{(2)}\left(z-z_n(1)\right)\delta^{(2)}\left(\omega-\frac{O_{nn}}{N}\right) \right\rangle_{X_1}.
\end{equation}
Note that such a definition implies the JPD normalization via the mean eigenvalue density:
\begin{equation}\label{norm}
\int_0^{\infty}{\cal P}_N(z_0,\omega)\,d\omega=\left\langle \rho_N(z_0)\right\rangle 
\end{equation}
and its associated first moment is simply the Chalker-Mehlig mean value
defined in \eqref{CMmean}:
\begin{equation}\label{1stmom}
\int_0^{\infty}{\cal P}_N(z_0,\omega)\,\omega\,d\omega={\cal O}_1(z_0). 
\end{equation}
Now recalling  \eqref{e1_r}, \eqref{e2a} and eventually \eqref{limit_nonorthB} one arrives at the identity 
\begin{eqnarray}\label{e1_r1}
\lim_{N\to \infty}\int_0^{\infty} \frac{1}{\pi \omega} e^{-\frac{|p|^2}{\omega}} {\cal P}_N(z_0,\omega)\,d\omega\nonumber 
\\=\left\langle\rho(z_0)\right\rangle^2\frac{2\left\langle{\cal O}(z_0)\right\rangle^2}{(\left\langle{\cal O}(z_0)\right\rangle+\pi\left\langle\rho(z_0)\right\rangle|p|^2)^3}.
\end{eqnarray}
The left-hand side of this relation is essentially the Laplace transform of the function $\frac{1}{\pi \omega}  {\cal P}_{\infty}(z,1/\omega)$, hence by straightforward Laplace inversion of the right-hand side with respect to Laplace variable $|p|^2$ we find
\begin{equation}\label{distrib_nonorth}
 {\cal P}_{\infty}(z,\omega)=\frac{1}{\pi^2\left\langle\rho(z_0)\right\rangle}\frac{\left\langle {\cal O}(z_0)\right\rangle^2}{\omega^3}e^{-\frac{\left\langle {\cal O}(z_0)\right\rangle}{\pi \left\langle\rho(z_0)\right\rangle \omega}}.
\end{equation}
 Note that  \eqref{distrib_nonorth}  implies
$\int_0^{\infty}{\cal P}_{\infty}(z_0,\omega)\,d\omega=\left\langle\rho(z_0)\right\rangle$, correctly matching the normalization in \eqref{norm}.  For the first moment we have $\int_0^{\infty}{\cal P}_{\infty}(z_0,\omega)\,\omega\,d\omega=\frac{1}{\pi}\left\langle {\cal O}(z_0)\right\rangle$, which by comparison with \eqref{1stmom} implies that 
$\left\langle {\cal O}(z_0)\right\rangle=\pi {\cal O}_1(z_0)$ as anticipated.
For the complex Ginibre case $\left\langle {\cal O}(z_0)\right\rangle=1-|z_0|^2$ for $|z_0|<1$ and \eqref{distrib_nonorth} coincides with the distribution for the diagonal non-orthogonality 
factor in the spectral bulk derived originally by approaches different from the present one in\cite{bourgadedoubach2020} and \cite{fyodorov2018CMP}.

As a comparison it is also useful to perform a similar computation for the case of complex symmetric matrices $X=X^T$, with due changes.
In such a case the left and right eigenvectors are related by transposition:
$\mathbf{l}_n^{\dagger}=\mathbf{r}_n^T$ and there is no sense in imposing the bi-orthogonality condition, so  $\mathbf{l}_n^{\dagger}\mathbf{r}_n\ne 1$.
To specify the Gaussian ensemble, we choose $X=\frac{U+iW}{\sqrt{2}}$, with $U,W$ being independent copies of real symmetric GOE matrices.
A straightforward computation shows that \eqref{e2a} should be replaced with \begin{equation}\label{e2asym}
F_{ab}(p)=\delta_{nn}\frac{N}{2\pi O_{nn}}e^{-\frac{N}{2O_{nn}}|p|^2}, 
\end{equation}
where $O_{nn}=\frac{\left(\mathbf{l}_n^{\dagger}\mathbf{l}_n\right)
\left(\mathbf{r}_n^{\dagger}\mathbf{r}_n\right)}{\left(\mathbf{l}_n^{\dagger}\mathbf{r}_n\right)
\left(\mathbf{r}_n^{\dagger}\mathbf{l}_n\right)}$ is the corresponding 
diagonal non-orthogonality factor. The limiting distribution ${\cal P}_{\infty}(z,\omega)$ of such factor for complex symmetric case has been found in \cite{akemann2025spectral} and is of the same form as in the complex Ginibre case, see \eqref{distrib_nonorth}, but with twice smaller value of $\left\langle {\cal O}(z_0)\right\rangle$, i.e. $\left\langle {\cal O}(z_0)\right\rangle=\frac{1}{2}(1-|z_0|^2)$.  Reversing the argument shows that the parametric correlations of eigenvalue densities in the complex symmetric matrices must necessarily start with exactly the same term as in \eqref{parcorden_lim2}, i.e. have the form $c_a(u)=\frac{2a^2}{(a+u)^3}+\cdots$, with exactly the same definition of $a$ and $u$. In turn, this implies that the parametric number covariance in both symmetry classes starts with the square-root term featuring in the first line of \eqref{parnummvar_result}.
Hence to see the difference in parametric number covariance between the two symmetry classes requires subtracting this common term. 


\onecolumngrid

\newpage
\beginsupplement

\begin{center}
\large\bfseries
Supplementary Material: Parametric correlations in dissipative quantum chaos: non-Hermitian random matrix approach
\end{center}

\section{Computation of the replica generating function}

The goal of this section is to detail how to compute the following replica partition function
\begin{align}\label{genfundef}
{\cal Z}_n(z_1,\overline{z_1},z_2, \overline{z_2})&=\left\langle \left|\det{z_1{\bf 1}_n-X_{1}}\right|^{2n} \left|\det{z_2{\bf 1}_n-X_{\alpha}}\right|^{2n}\right \rangle \;, \\
{\rm where}\;\;X_{\alpha}&=\alpha X_1+\sqrt{1-\alpha^2}X_2\;,
\end{align} 
$0\leq \alpha\leq 1$ and $X_{1,2}$ are independent complex Ginibre matrices. Note that one can alternatively take $\alpha\in \mathbb{C}$ but the only relevant parameter is $|\alpha|$ due to the invariance of the complex Ginibre ensemble. To compute this term, we rewrite the expression as
\begin{align}
{\cal Z}_n=&\left\langle \det\begin{pmatrix}
    {\bf 0}&{\bf 0}&i(z_1{\bf 1}_N-X_1)&{\bf 0}\\
    {\bf 0}&{\bf 0}&{\bf 0}&i(z_2{\bf 1}_N-X_\alpha)\\
    i(\bar z_1{\bf 1}_N-X_1^{\dagger})&{\bf 0}&{\bf 0}&{\bf 0}\\
    {\bf 0}&i(\bar z_2{\bf 1}_N-X_\alpha^{\dagger})&{\bf 0}&{\bf 0}
\end{pmatrix}^n\right \rangle_{X_1,X_2}\\
=&\int {\cal D}(\varphi,\psi,\eta,\xi)\left\langle  e^{i\sum_{a=1}^n \begin{pmatrix}
    \varphi_{1a}^{\dagger}&\eta_{1a}^{\dagger}&\varphi_{2a}^{\dagger}&\eta_{2a}^{\dagger}
\end{pmatrix}\begin{pmatrix}
    {\bf 0}&{\bf 0}&z_1{\bf 1}_N-X_1&{\bf 0}\\
    {\bf 0}&{\bf 0}&{\bf 0}&z_2{\bf 1}_N-X_\alpha\\
    \bar z_1{\bf 1}_N-X_1^{\dagger}&{\bf 0}&{\bf 0}&{\bf 0}\\
    {\bf 0}&\bar z_2{\bf 1}_N-X_\alpha^{\dagger}&{\bf 0}&{\bf 0}
\end{pmatrix}\begin{pmatrix}
    \psi_{1a}\\\xi_{1a}\\\psi_{2a}\\\xi_{2a}
\end{pmatrix}}\right \rangle_{X_1,X_2}\nonumber\\
=&\int {\cal D}(\varphi,\psi,\eta,\xi)e^{i\sum_{a=1}^n \left[z_1\varphi_{1a}^{\dagger}\psi_{2a}+\bar z_1 \varphi_{2a}^{\dagger}\psi_{1a}+z_2\eta_{1a}^{\dagger}\xi_{2a}+\bar z_2 \eta_{2a}^{\dagger}\xi_{1a}\right]}\left\langle  e^{i\sqrt{1-\alpha^2}\Tr\left[X_2\sum_{a=1}^n\xi_{2a}\otimes \eta_{1a}^{\dagger}+X_2^{\dagger}\sum_{a=1}^n\xi_{1a}\otimes \eta_{2a}^{\dagger}\right]}\right \rangle_{X_2}\nonumber\\
&\times \left\langle  e^{i\Tr\left[X_1\sum_{a=1}^n\left(\psi_{2a}\otimes \varphi_{1a}^{\dagger}+\alpha \xi_{2a}\otimes \eta_{1a}^{\dagger}\right)+X_1^{\dagger}\sum_{a=1}^n\left(\psi_{1a}\otimes \varphi_{2a}^{\dagger}+\alpha \xi_{1a}\otimes \eta_{2a}^{\dagger}\right)\right]}\right \rangle_{X_1}\nonumber\;.
\end{align} 
Next we use the following identity
\begin{align}
    \left\langle  \exp\left(i c \Tr\left[XA+X^{\dagger}B\right]\right) \right\rangle_{X}=\exp\left(-\frac{c^2}{N}\Tr\left[AB\right]\right)
\end{align}
Using this identity, one obtains 
\begin{align}
{\cal Z}_n=&\int {\cal D}(\varphi,\psi,\eta,\xi)e^{i\sum_{a=1}^n \left[z_1\varphi_{1a}^{\dagger}\psi_{2a}+\bar z_1 \varphi_{2a}^{\dagger}\psi_{1a}+z_2\eta_{1a}^{\dagger}\xi_{2a}+\bar z_2 \eta_{2a}^{\dagger}\xi_{1a}\right]}\\
&\times e^{\frac{1}{N}\sum_{a,b}\left[(\varphi_{1a}^{\dagger}\psi_{1b})(\varphi_{2b}^{\dagger}\psi_{2a})+(\eta_{1a}^{\dagger}\xi_{1b})(\eta_{2b}^{\dagger}\xi_{2a})+\alpha\left[(\eta_{1a}^{\dagger}\psi_{1b})(\varphi_{2b}^{\dagger}\xi_{2a})+(\varphi_{1a}^{\dagger}\xi_{1b})(\eta_{2b}^{\dagger}\psi_{2a})\right]\right]}\nonumber\;.
\end{align}
Next, we use the Hubbard-Stratonovic formulae
\begin{align}
 &e^{\frac{1}{N}\sum_{a,b}\left[(\varphi_{1a}^{\dagger}\psi_{1b})(\varphi_{2b}^{\dagger}\psi_{2a})+(\eta_{1a}^{\dagger}\xi_{1b})(\eta_{2b}^{\dagger}\xi_{2a})+\alpha\left[(\eta_{1a}^{\dagger}\psi_{1b})(\varphi_{2b}^{\dagger}\xi_{2a})+(\varphi_{1a}^{\dagger}\xi_{1b})(\eta_{2b}^{\dagger}\psi_{2a})\right]\right]}=\left(\frac{N}{2\pi}\right)^{4n^2}\int dQ dQ^{\dagger} e^{-N\Tr(Q Q^{\dagger})}\\
 &\times e^{-\sum_{a,b}\left[q_{ab}^{11}(\varphi_{1a}^{\dagger}\psi_{1b})+\bar q_{ab}^{11}(\varphi_{2b}^{\dagger}\psi_{2a})+q_{ab}^{22}(\eta_{1a}^{\dagger}\xi_{1b})+\bar q_{ab}^{22}(\eta_{2b}^{\dagger}\xi_{2a})+\sqrt{\alpha}\left[q_{ab}^{21}(\eta_{1a}^{\dagger}\psi_{1b})+\bar q_{ab}^{21}(\varphi_{2b}^{\dagger}\xi_{2a})+q_{ab}^{12}(\varphi_{1a}^{\dagger}\xi_{1b})+\bar q_{ab}^{12}(\eta_{2b}^{\dagger}\psi_{2a})\right]\right]}\nonumber\;,\\
 &Q_{\alpha}=\begin{pmatrix}
 q^{11}&\sqrt{\alpha}q^{12}\\
 \sqrt{\alpha}q^{21}&q^{22}
 \end{pmatrix}\;,\;\;Q\equiv Q_{\alpha=1}\;.
\end{align}
Introducing these integrals, one can compute explicitly the determinant as
\begin{align}
{\cal Z}_n=&\left(\frac{N}{2\pi}\right)^{4n^2}\int dQ dQ^{\dagger} e^{-N\Tr(Q Q^{\dagger})}\int {\cal D}(\varphi,\psi,\eta,\xi)e^{-\sum_{a,b=1}^n \begin{pmatrix}
    \varphi_{1a}^{\dagger}&\eta_{1a}^{\dagger}&\varphi_{2a}^{\dagger}&\eta_{2a}^{\dagger}
\end{pmatrix}\begin{pmatrix}
    Q_{\alpha}&-i\begin{pmatrix}
    z_1{\bf 1}_n&0\\
    0&z_2{\bf 1}_n
    \end{pmatrix}\\
    -i\begin{pmatrix}
    \bar z_1{\bf 1}_n&0\\
    0&\bar z_2{\bf 1}_n
    \end{pmatrix}&Q_{\alpha}^{\dagger}
\end{pmatrix}\begin{pmatrix}
    \psi_{1b}\\\xi_{1b}\\\psi_{2b}\\\xi_{2b}
\end{pmatrix}}\nonumber\\
=&\left(\frac{N}{2\pi}\right)^{4n^2}\int dQ dQ^{\dagger} \exp\left(-N\Tr\left[Q Q^{\dagger}-\ln\det\begin{pmatrix}
    Q_{\alpha}&-i\begin{pmatrix}
    z_1{\bf 1}_n&0\\
    0&z_2{\bf 1}_n
    \end{pmatrix}\\
    -i\begin{pmatrix}
    \bar z_1{\bf 1}_n&0\\
    0&\bar z_2{\bf 1}_n
    \end{pmatrix}&Q_{\alpha}^{\dagger}
\end{pmatrix}\right]\right)\;.
\end{align}
The expression above is exact. Next, we will analyse this formula in a specific scaling regime.

\section{Scaling regime}

We now consider the following specific scaling regime, $z_{1,2}\to z$ and $\alpha\to 1$ as $N\to \infty$ with the following finite parameters
\begin{equation}
    v^2=2N(1-\alpha)=O(1)\;,\;\;\omega=2\sqrt{N}(z_1-z)=2\sqrt{N}(z-z_2)=O(1)\;.
\end{equation}
Introducing the matrices
\begin{equation}
\tilde Q=\begin{pmatrix}
 0&q^{12}\\
 q^{21}&0
 \end{pmatrix}\;,\;\;\Lambda=\begin{pmatrix}
 {\bf 1}_n&0\\
 0&-{\bf 1}_n
 \end{pmatrix}\;,    
\end{equation}
we can expand the determinant as $N\gg 1$, yielding
\begin{align}
    &N\ln \det\begin{pmatrix}
    Q_{\alpha}&-i\begin{pmatrix}
    z_1{\bf 1}_n&0\\
    0&z_2{\bf 1}_n
    \end{pmatrix}\\
    -i\begin{pmatrix}
    \bar z_1{\bf 1}_n&0\\
    0&\bar z_2{\bf 1}_n
    \end{pmatrix}&Q_{\alpha}^{\dagger}
\end{pmatrix}=N\ln \det\begin{pmatrix}
    Q&-i z{\bf 1}_{2n}\\
    -i\bar z{\bf 1}_{2n}&Q^{\dagger}
\end{pmatrix}\\
&+N\Tr\ln\left({\bf 1}_{4n}-\frac{1}{N}\begin{pmatrix}
    Q&-i z{\bf 1}_{2n}\\
    -i\bar z{\bf 1}_{2n}&Q^{\dagger}
\end{pmatrix}^{-1}\begin{pmatrix}
    \frac{v^2}{4}\tilde Q&-\frac{i\sqrt{N}\omega}{2}\Lambda\\
    -\frac{i\sqrt{N}\bar \omega}{2}\Lambda&\frac{v^2}{4}\tilde Q^{\dagger}
\end{pmatrix}\right)\nonumber\\
&=N\ln \det(QQ^{\dagger}+|z|^2{\bf 1}_{2n})+N\Tr\ln\left({\bf 1}_{4n}-\frac{1}{4N}\begin{pmatrix}
    Q^{\dagger}(Q Q^{\dagger}+|z|^2{\bf 1}_{2n})^{-1}&i z(Q^{\dagger} Q+|z|^2{\bf 1}_{2n})^{-1}\\
    i\bar z(Q Q^{\dagger}+|z|^2{\bf 1}_{2n})^{-1}&Q(Q^{\dagger} Q+|z|^2{\bf 1}_{2n})^{-1}
\end{pmatrix}\begin{pmatrix}
    v^2\tilde Q&-2i\sqrt{N}\omega\Lambda\\
    -2i\sqrt{N}\bar \omega\Lambda&v^2\tilde Q^{\dagger}
\end{pmatrix}\right)\nonumber
\end{align}
Next, we introduce the singular value decomposition $Q=U R^{1/2}V$ and $Q^{\dagger}=V^{\dagger}R^{1/2}U^{\dagger}$ with $UU^{\dagger}=VV^{\dagger}={\bf 1}_{2n}$. Using this decomposition, one can express $Q^{\dagger}(Q Q^{\dagger}+|z|^2{\bf 1}_{2n})^{-1}=V^{\dagger}R^{1/2}(R+|z|^2{\bf 1}_{2n})^{-1}U^{\dagger}$ and $Q(Q^{\dagger} Q+|z|^2{\bf 1}_{2n})^{-1}=UR^{1/2}(R+|z|^2{\bf 1}_{2n})^{-1}V$. The integration measure takes the form
\begin{equation}
    dQdQ^{\dagger}=C_n d\mu(U)d\mu(V)\prod_{a<b}^{2n}|r_a-r_b|^2\prod_{a=1}^{2n} dr_a\;.
\end{equation}
The (exponential) leading order is obtained by considering the solution $r_a=r_*$ and we expand around this solution as $r_a=r_*+\zeta_a/\sqrt{N}$, such that
\begin{align}
    &N\left[-\Tr(QQ^{\dagger})+\ln \det(QQ^{\dagger}+|z|^2{\bf 1}_{2n})\right]=N\left[-\Tr(R)+\ln \det(R+|z|^2{\bf 1}_{2n})\right]\\
    &=2Nn\left[-r_*+\ln(r_*+|z|^2)\right]+\sum_{a}\left(-1+\frac{1}{r_*+|z|^2}\right)\sqrt{N}\zeta_a-\frac{1}{2(r_*+|z|^2)^2}\sum_a\zeta_a^2+o(1)\nonumber\\
    &=-2Nn(1-|z|^2)-\frac{1}{2}\sum_a\zeta_a^2+o(1)\nonumber
\end{align}
where we have used the solution $r_*=1-|z|^2\geq 0$. The associated integration measure is then
\begin{equation}
    dQdQ^{\dagger}=\frac{C_n}{N^{2n^2}} d\mu(U)d\mu(V)\prod_{a<b}^{2n}|\zeta_a-\zeta_b|^2\prod_{a=1}^{2n} d\zeta_a\;.
\end{equation}
Next, we expand the matrix 
\begin{align}
 &\begin{pmatrix}
    Q^{\dagger}(Q Q^{\dagger}+|z|^2{\bf 1}_{2n})^{-1}&i z(Q^{\dagger} Q+|z|^2{\bf 1}_{2n})^{-1}\\
    i\bar z(Q Q^{\dagger}+|z|^2{\bf 1}_{2n})^{-1}&Q(Q^{\dagger} Q+|z|^2{\bf 1}_{2n})^{-1}
\end{pmatrix}\\
&=\begin{pmatrix}
    \sqrt{r_*}V^{\dagger}U^{\dagger}&i z\\
    i\bar z&\sqrt{r_*}UV
\end{pmatrix}+\frac{1}{\sqrt{N}}\begin{pmatrix}
    \left(\frac{1}{2\sqrt{r_*}}-\sqrt{r_*}\right)V^{\dagger}\zeta U^{\dagger}&-i z V^{\dagger}\zeta V\\
    -i\bar z U\zeta U^{\dagger}&\left(\frac{1}{2\sqrt{r_*}}-\sqrt{r_*}\right)U\zeta V
\end{pmatrix}
\end{align}
Thus, the product of matrices reads
\begin{align}
    &\frac{1}{4N}\begin{pmatrix}
    Q^{\dagger}(Q Q^{\dagger}+|z|^2{\bf 1}_{2n})^{-1}&i z(Q^{\dagger} Q+|z|^2{\bf 1}_{2n})^{-1}\\
    i\bar z(Q Q^{\dagger}+|z|^2{\bf 1}_{2n})^{-1}&Q(Q^{\dagger} Q+|z|^2{\bf 1}_{2n})^{-1}
\end{pmatrix}\begin{pmatrix}
    v^2\tilde Q&-2i\sqrt{N}\omega\Lambda\\
    -2i\sqrt{N}\bar \omega\Lambda&v^2\tilde Q^{\dagger}
\end{pmatrix}=\frac{1}{2\sqrt{N}}A+\frac{1}{4N}B\;,\\
A&=\begin{pmatrix}
    z\bar \omega\Lambda&-i\omega\sqrt{r_*}V^{\dagger}U^{\dagger}\Lambda\\
    -i\bar \omega\sqrt{r_*}UV\Lambda&\bar z \omega\Lambda
\end{pmatrix}\;,\nonumber\\
B&=\begin{pmatrix}
    v^2\sqrt{r_*}V^{\dagger}U^{\dagger}\tilde Q-2\bar \omega z V^{\dagger}\zeta V\Lambda&i\left[z\tilde Q^{\dagger}-\omega\left(\frac{1}{\sqrt{r_*}}-2\sqrt{r_*}\right)V^{\dagger}\zeta U^{\dagger}\Lambda\right]\\
    i\left[\bar z\tilde Q-\bar \omega\left(\frac{1}{\sqrt{r_*}}-2\sqrt{r_*}\right)U\zeta V\Lambda\right]&v^2\sqrt{r_*}UV\tilde Q^{\dagger}-2\bar z\omega U\zeta U^{\dagger}\Lambda
\end{pmatrix}\;.\nonumber
\end{align}
Expanding the following term, one obtains
\begin{align}
    &N\Tr\ln\left({\bf 1}_{4n}-\frac{1}{2\sqrt{N}}A-\frac{1}{4N}B\right)=-\frac{\sqrt{N}}{2}\Tr(A)-\frac{1}{4}\Tr\left(B+\frac{A^2}{2}\right)+o(1)\\
    &=-\frac{1}{4}\Tr[v^2\sqrt{r_*}(V^{\dagger}U^{\dagger}\tilde Q+UV\tilde Q^{\dagger})-2(\bar \omega z V^{\dagger}\zeta V+\bar z\omega U\zeta U^{\dagger})\Lambda]-\frac{1}{8}\Tr\left[[(z\bar \omega)^2+(\bar z\omega)^2]{\bf 1}_{2n}-2|\omega|^2r_*UV\Lambda V^{\dagger}U^{\dagger}\Lambda\right]+o(1)\nonumber
\end{align}
Note that the matrix $A$ is traceless, thus the leading $\sqrt{N}$ term vanishes. Let us denote $P=V\zeta V^{\dagger}$ and $U_0=UV$. The terms above can be rewritten as
\begin{equation}
\frac{1}{2}\Tr(P(\bar \omega z \Lambda+\bar z\omega U_0^{\dagger}\Lambda U_0))-\frac{v^2\sqrt{r_*}}{4}\Tr[U_0^{\dagger}\tilde Q+U_0\tilde Q^{\dagger}]-\frac{n}{4}[(z\bar \omega)^2+(\bar z\omega)^2]+\frac{|\omega|^2r_*}{4}\Tr\left[U_0\Lambda U_0^{\dagger}\Lambda\right]    
\end{equation}
Next, we rewrite
\begin{equation}
    dPd\mu(U_0)=K_n d\mu(V)d\mu(U)\prod_{a<b}^{2n}|\zeta_a-\zeta_b|^2\prod_{a=1}^{2n} d\zeta_a\;.
\end{equation}
Using the identity $\Tr(P^2)=\Tr(\zeta^2)=\sum_a\zeta_a^2$, this allows in particular to express
\begin{align}
{\cal Z}_n\propto&\int d^2P d\mu(U_0) e^{-2Nn(1-|z|^2)-\frac{1}{2}\Tr(P^2)+\frac{1}{2}\Tr(P(\bar \omega z \Lambda+\bar z\omega U_0^{\dagger}\Lambda U_0))-\frac{v^2\sqrt{r_*}}{4}\Tr[U_0^{\dagger}\tilde Q+U_0\tilde Q^{\dagger}]-\frac{n}{4}[(z\bar \omega)^2+(\bar z\omega)^2]+\frac{|\omega|^2r_*}{4}\Tr\left[U_0\Lambda U_0^{\dagger}\Lambda\right]}\nonumber\\
\propto&\int d\mu(U_0) e^{-2Nn(1-|z|^2)+\frac{1}{8}\Tr[(\bar \omega z \Lambda+\bar z\omega \Lambda U_0^{\dagger}\Lambda U_0)^2]-\frac{v^2\sqrt{r_*}}{4}\Tr[U_0^{\dagger}\tilde Q+U_0\tilde Q^{\dagger}]-\frac{n}{4}[(z\bar \omega)^2+(\bar z\omega)^2]+\frac{|\omega|^2r_*}{4}\Tr\left[U_0\Lambda U_0^{\dagger}\Lambda\right]}\nonumber\\
\propto&\int d\mu(U_0) e^{-2Nn(1-|z|^2)+\frac{n}{4}[(\bar \omega z)^2+(\omega \bar z)^2]+\frac{|z|^2|\omega|^2}{4}\Tr[\Lambda U_0^{\dagger}\Lambda U_0]-\frac{v^2\sqrt{(1-|z|^2)}}{4}\Tr[U_0^{\dagger}\tilde Q+U_0\tilde Q^{\dagger}]-\frac{n}{4}[(z\bar \omega)^2+(\bar z\omega)^2]+\frac{|\omega|^2(1-|z|^2)}{4}\Tr\left[\Lambda U_0^{\dagger}\Lambda U_0\right]}\nonumber\\
\propto&\int d\mu(U_0) e^{-2Nn(1-|z|^2)-\frac{v^2\sqrt{(1-|z|^2)}}{4}\Tr[U_0^{\dagger}\tilde Q+U_0\tilde Q^{\dagger}]+\frac{|\omega|^2}{4}\Tr\left[\Lambda U_0^{\dagger}\Lambda U_0\right]}\;.
\end{align}
To simplify this expression, we use that at leading order
\begin{equation}
Q=UR_*^{1/2}V=\sqrt{r_*}UV=\sqrt{r_*}U_0=\sqrt{r_*}\begin{pmatrix}
        u_0^{11}&u_0^{12}\\
        u_0^{21}&u_0^{22}
    \end{pmatrix}\;,\;\;\tilde Q=\sqrt{r_*}\begin{pmatrix}
        0&u_0^{12}\\
        u_0^{21}&0
    \end{pmatrix}\;.
\end{equation}
One can rewrite the terms
\begin{align}
\Tr[U_0^{\dagger}\tilde Q+U_0\tilde Q^{\dagger}]&=\sqrt{r_*}\Tr\left[\begin{pmatrix}
        {u_0^{11}}^{\dagger}&{u_0^{21}}^{\dagger}\\
        {u_0^{12}}^{\dagger}&{u_0^{22}}^{\dagger}
    \end{pmatrix}\begin{pmatrix}
        0&u_0^{12}\\
        u_0^{21}&0
    \end{pmatrix}+\begin{pmatrix}
        u_0^{11}&u_0^{12}\\
        u_0^{21}&u_0^{22}
    \end{pmatrix}\begin{pmatrix}
        0&{u_0^{21}}^{\dagger}\\
        {u_0^{12}}^{\dagger}&0
    \end{pmatrix}\right]\\
    &=2\sqrt{r_*}\Tr\left[u_0^{12}{u_0^{12}}^{\dagger}+u_0^{21}{u_0^{21}}^{\dagger}\right]\;,\nonumber\\
    \Tr\left[\Lambda U_0^{\dagger}\Lambda U_0\right]&=\Tr\left[\begin{pmatrix}
        {\bf 1}_{n}&0\\
        0&-{\bf 1}_{n}
    \end{pmatrix}\begin{pmatrix}
        {u_0^{11}}^{\dagger}&{u_0^{21}}^{\dagger}\\
        {u_0^{12}}^{\dagger}&{u_0^{22}}^{\dagger}
    \end{pmatrix}\begin{pmatrix}
        {\bf 1}_{n}&0\\
        0&-{\bf 1}_{n}
    \end{pmatrix}\begin{pmatrix}
        u_0^{11}&u_0^{12}\\
        u_0^{21}&u_0^{22}
    \end{pmatrix}\right]\\
    &=\Tr\left[{u_0^{11}}^{\dagger}u_0^{11}-{u_0^{21}}^{\dagger}u_0^{21}+{u_0^{22}}^{\dagger}u_0^{22}-{u_0^{12}}^{\dagger}u_0^{12}\right]\nonumber\\
    &=2n-2\Tr\left[{u_0^{12}}^{\dagger}u_0^{12}+{u_0^{21}}^{\dagger}u_0^{21}\right]\;,\nonumber
\end{align}
where we have used that ${u_0^{11}}^{\dagger}u_0^{11}+{u_0^{21}}^{\dagger}u_0^{21}={\bf 1}_{n}$ and ${u_0^{22}}^{\dagger}u_0^{22}+{u_0^{12}}^{\dagger}u_0^{12}={\bf 1}_{n}$. This yields
\begin{equation}
{\cal Z}_n\propto e^{-2n(1-|z|^2)\left(N+\frac{v^2}{4}\right)}\int d\mu(U_0) e^{\frac{|\omega|^2+v^2(1-|z|^2)}{4}\Tr\left[\Lambda U_0^{\dagger}\Lambda U_0\right]}\;.  
\end{equation}
This is our final expression for the replicated generating function. In the next section we analyse how to obtain from this expression the two point density correlation function. 

\section{Correlation functions}
To obtain the expression of the two point density correlation function, we denote $z_1=z+w/2$ and $z_2=z-w/2$ and need to compute
\begin{equation}\nonumber
\left\langle \rho_{1}(z_1)\rho_{\alpha}(z_2)\right\rangle=\lim_{n\to 0}\frac{1}{\pi n^2}
\frac{\partial}{\partial z_2} \frac{\partial}{\partial z_1} \frac{\partial}{\partial \overline{z}_2} \frac{\partial}{\partial \overline{z}_1}   {\cal Z}_n(z_1,\overline{z_1},z_2, \overline{z_2})=\lim_{n\to 0}\frac{1}{\pi n^2}
\left(\frac{1}{4}\frac{\partial^2}{\partial z^2}-\frac{\partial^2}{\partial w^2}\right)\left(\frac{1}{4}\frac{\partial^2}{\partial \bar z^2}-\frac{\partial^2}{\partial \bar w^2}\right){\cal Z}_n\;.
\end{equation}
Using that at leading order
\begin{align}
{\cal Z}_n&\approx \phi_n(N z\bar z)\psi_n(Nw\bar w)\;,\;\;\phi_n(x)=e^{-2n(N- x)}\;,\\
    \psi_n\left(y\right)&=e^{\frac{n}{2}\left(y+v^2 r_*\right)-n^2b(y)+O(n)^3}\;,\;\;b(y)=\ln(y+v^2 r_*)+\delta E(y+v^2 r_*)\;,   
\end{align}
where $\delta E(x)=\int_1^{\infty}dt\,e^{-xt}(t-1)/t^2$. 
One obtains at leading order
\begin{align}
\left\langle \rho_{1}(z_1)\rho_{\alpha}(z_2)\right\rangle=&\lim_{n\to 0}\left[-\frac{N^4}{4\pi n^2}(\bar w^2z^2+w^2\bar z^2)\psi_n''\left(Nw\bar w\right)\phi_n''(N z\bar z)\right.\\
&+\frac{N^2}{16\pi n^2}\psi_n(N z\bar z)\left(2\phi_n''(N z\bar z)+4Nz\bar z\phi_n^{(3)}(N z\bar z)+(Nz\bar z)^2\phi_n^{(4)}(N z\bar z)\right)\nonumber\\
&\left.+\frac{N^2}{\pi n^2}\phi_n(N z\bar z)\left(2\psi_n''(N w\bar w)+4Nw\bar w\psi_n^{(3)}(N w\bar w)+(Nw\bar w)^2\psi_n^{(4)}(N w\bar w)\right)\right]\nonumber\\
=&\frac{N^2}{\pi}\left[1-2b''(N w\bar w)-4(Nw\bar w) b^{(3)}(N w\bar w)-(Nw\bar w)^2 b^{(4)}(N w\bar w)\right]\;.\nonumber
\end{align}
Introducing $w=\omega/\sqrt{N}$ and $u=\omega\bar \omega$, one can express the connected two-point density correlation function in terms of the scaling form
\begin{align}
 &\left\langle \rho_{1}(z_1)\rho_{\alpha}(z_2)\right\rangle-\left\langle \rho_{1}(z_1)\right\rangle\left\langle\rho_{\alpha}(z_2)\right\rangle= \frac{1}{\pi^2}c_{a=v^2 (1-|z|^2)}\left(u=\omega\bar \omega\right)\;, \\
 c_a(u)=&-e^{-(a+u)}-\frac{2a}{(a+u)^3}\left(2(1-e^{-(a+u)})-(a+u)(2+a+u)e^{-(a+u)}\right)\\
 &+\frac{a^2}{(a+u)^4}\left(6(1-e^{-(a+u)})-(a+u)(4+a+u)e^{-(a+u)}\right)\;.
\end{align}
It will be useful to use the following identities
\begin{align}
  &\frac{1}{x^3}\left(2(1-e^{-x})-x(2+x)e^{-x}\right)=\partial_x^2\left(\frac{1-e^{-x}}{x}\right)\;,\\
  &\frac{1}{x^4}\left(6(1-e^{-x})-x(4+x)e^{-x}\right)=\frac{2}{x^3}-\left[\partial_x^2+\partial_x^3\right]\left(\frac{1-e^{-x}}{x}\right)\;.
\end{align}
Thus, we can rewrite the scaling function
\begin{align}
   c_a(u)&=\frac{2a^2}{(a+u)^3}-e^{-(a+u)}-\left[a(2+a)\frac{\partial^2}{\partial a^2}+a^2\frac{\partial^3}{\partial a^3}\right]\frac{1-e^{-(a+u)}}{a+u}\;. \nonumber
\end{align}
Using the expression of $c_a(u)$, one can easily check that
\begin{equation}
 \lim_{v\to 0} v^2 c_{v^2 (1-|z|^2)}\left(v^2|p|^2\right)=\frac{2(1-|z|^2)^2}{\left(|p|^2+1-|z|^2\right)^3}\;.   
\end{equation}

Next, we compute the correlation function between the number of eigenvalues. To compute this integral, we rely on the following identity
\begin{align}
    &N^2\int_{|z_1-z_0|<\frac{R}{\sqrt{N}}} d^2 z_1 \int_{|z_2-z_0|<\frac{R}{\sqrt{N}}} d^2 z_2\,F_a(\pi N\langle\rho(z_0)\rangle|z_1-z_2|^2)\\
    &=2N^2\pi\int_0^{\frac{R}{\sqrt{N}}}r_1 dr_1\int_0^{\frac{R}{\sqrt{N}}}r_2 dr_2\int_0^{2\pi}d\theta\,F_a\left[\pi N\langle\rho(z_0)\rangle(r_1^2+r_2^2-2r_1 r_2 \cos(\theta))\right]\nonumber\\
    &=\frac{\pi}{2(\pi \langle\rho(z_0)\rangle)^2}\int_0^{\pi \langle\rho(z_0)\rangle R^2} du_1\int_0^{\pi \langle\rho(z_0)\rangle R^2}du_2\int_0^{2\pi}d\theta\,F_a\left[u_1+u_2-2\sqrt{u_1u_2}\cos(\theta)\right]\nonumber\\
    &=\frac{\pi}{2(\pi \langle\rho(z_0)\rangle)^2}\int_0^{\infty}dx\,F_a\left(x\right)\int_0^{\pi \langle\rho(z_0)\rangle R^2} du_1\int_0^{\pi \langle\rho(z_0)\rangle R^2}du_2\int_0^{2\pi}d\theta\,\delta(u_1+u_2-2\sqrt{u_1u_2}\cos(\theta)-x)\nonumber\\
    &=\frac{\pi}{2(\pi \langle\rho(z_0)\rangle)^2}\int_0^{\infty}dx\,x\,F_a\left(x\right)\int_0^{\frac{\pi \langle\rho(z_0)\rangle R^2}{x}} dv_1\int_0^{\frac{\pi \langle\rho(z_0)\rangle R^2}{x}}dv_2\int_0^{2\pi}d\theta\,\delta(v_1+v_2-2\sqrt{v_1v_2}\cos(\theta)-1)\nonumber\\
&=\frac{\pi}{2(\pi \langle\rho(z_0)\rangle)^2}\int_0^{\infty}dx\,x\,F_a\left(x\right)\Psi\left(\frac{\pi \langle\rho(z_0)\rangle R^2}{x}\right)\;.\nonumber
\end{align}
It will be convenient to rewrite the following integral
\begin{align}
    \Psi(t)&=\int_0^{t} dv_1\int_0^{t}dv_2\int_0^{2\pi}d\theta\,\delta(v_1+v_2-2\sqrt{v_1v_2}\cos(\theta)-1)=\frac{1}{2}\int_0^{t} \frac{dv_1}{\sqrt{v_1}}\int_0^{t}\frac{dv_2}{\sqrt{v_2}}\int_0^{2\pi}d\theta\,\delta\left(\frac{v_1+v_2-1}{2\sqrt{v_1v_2}}-\cos(\theta)\right)\nonumber\\
    &=2\int_0^{t}dv_1\int_0^{t}dv_2\frac{\Theta\left(4v_1v_2-(v_1+v_2-1)^2\right)}{\sqrt{4v_1v_2-(v_1+v_2-1)^2}}\nonumber\\
    &=2\int_0^{t}dv_1\int_0^{t}dv_2\frac{\Theta\left((v_1+v_2)^2-(v_1-v_2)^2-(v_1+v_2-1)^2\right)}{\sqrt{2(v_1+v_2)-(v_1-v_2)^2-1}}\nonumber\\
    &=4\int_0^{\frac{t}{2}} du\,\int_{0}^{2u} dr\,\frac{\Theta\left(4u-1-r^2\right)}{\sqrt{4u-1-r^2}}+4\int_{\frac{t}{2}}^{t} du\,\int_{0}^{2(t-u)} dr\,\frac{\Theta\left(4u-1-r^2\right)}{\sqrt{4u-1-r^2}}\;.
\end{align}
Taking the derivative of this expression with respect to $t$, one can check that the only remaining term reads
\begin{align}
    \Psi'(t)&=8\Theta(4t-1)\int_{\frac{t}{2}}^{t} du\,\frac{\Theta\left(4u-1-4(t-u)^2\right)}{\sqrt{4u-1-4(t-u)^2}}=8\Theta(4t-1)\int_{\frac{t}{2}}^{t} du\,\frac{\Theta\left(4u-1-4(t-u)^2\right)}{\sqrt{4t-(2(t-u)+1)^2}}\\
    &=4\Theta(4t-1)\int_{\frac{1}{2\sqrt{t}}}^{\frac{1}{2}\left(\sqrt{t}+\frac{1}{\sqrt{t}}\right)^2} du\,\frac{\Theta\left(1-w^2\right)}{\sqrt{1-w^2}}=4\Theta(4t-1)\int_{\frac{1}{2\sqrt{t}}}^{1} du\,\frac{1}{\sqrt{1-w^2}}=4\Theta(4t-1){\rm arctan}\left(\sqrt{4t-1}\right)\;.\nonumber
\end{align}
Integrating with respect to $t$, one obtains
\begin{equation}
     \Psi(t)=4\Theta(4t-1)\left[t\,{\rm arctan}\left(\sqrt{4t-1}\right)-\frac{\displaystyle \sqrt{4t-1}}{\displaystyle 4}\right]\;.
\end{equation}
Inserting $t=(2\cos(\varphi/2))^{-2}$, one can show that
\begin{equation}
     \frac{1}{t}\Psi(t)=2(\varphi-\sin(\varphi))
\end{equation}
Thus, applying the change of variables $x=4\pi \langle\rho(z_0)\rangle R^2\cos(\varphi/2)^2=2\pi \langle\rho(z_0)\rangle R^2(1+\cos(\varphi))$, one obtains
\begin{align}
    &\frac{\pi}{2(\pi \langle\rho(z_0)\rangle)^2}\int_0^{\infty}dx\,x\,F_a\left(x\right)\Psi\left(\frac{\pi \langle\rho(z_0)\rangle R^2}{x}\right)\\
    &=\frac{\pi R^2}{2\pi \langle\rho(z_0)\rangle}\int_0^{4\pi \langle\rho(z_0)\rangle R^2}dx\,F_a\left(x\right)\left(\frac{\pi \langle\rho(z_0)\rangle R^2}{x}\right)^{-1}\Psi\left(\frac{\pi \langle\rho(z_0)\rangle R^2}{x}\right)\\
    &=2\pi R^4 \int_0^{\pi}d\varphi \sin(\varphi)[\varphi-\sin(\varphi)]F_a\left(2\pi \langle\rho(z_0)\rangle R^2(1+\cos(\varphi))\right) \;.
\end{align}
Inserting $F_{a}(x)=-\langle\rho(z_0)\rangle^2 e^{-(a+x)}$ and denoting $y=\pi \langle\rho(z_0)\rangle R^2$ in the equation above, one obtains
\begin{align}
    &-\frac{2}{\pi}y^2\int_0^{\pi}d\varphi \sin(\varphi)[\varphi-\sin(\varphi)]e^{-a-2y(1+\cos(\varphi))}\nonumber\\
    &=-\frac{y}{\pi}\left[(\varphi-\sin(\varphi))e^{-a-2y(1+\cos(\varphi))}\right]_0^{\pi}+\frac{y}{\pi} e^{-a-2y}\int_0^{\pi}d\varphi [1-\cos(\varphi)]e^{-2y\cos(\varphi)}\nonumber\\
    &=y e^{-a}\left(e^{-2y}\left[I_0\left(2y\right)+I_1\left(2y\right)\right]-1\right)\;.
\end{align}
Next, we compute the following term in the expression of $F_{a}(x)=2\langle\rho(z_0)\rangle^2 a^2/(a+x)^3$, which reads 
\begin{align}
    &\frac{4 a^2 y^2}{\pi}\int_0^{\pi}d\varphi \frac{\sin(\varphi)[\varphi-\sin(\varphi)]}{(a+2y(1+\cos(\varphi)))^3}=\frac{a^2 y}{\pi}\left[\frac{\varphi-\sin(\varphi)}{(a+2y(1+\cos(\varphi)))^2}\right]_0^{\pi}-\frac{a^2 y}{\pi}\int_0^{\pi}d\varphi \frac{1-\cos(\varphi)}{(a+2y(1+\cos(\varphi)))^2}\nonumber\\
    =&y-\frac{a^2 }{\pi}\int_0^{\frac{\pi}{2}}d\phi \frac{4y\sin(\phi)^2}{(a+4y\cos(\phi)^2)^2}=y\left[1-\sqrt{\frac{a}{a+4y}}\right]
\end{align}
Finally, inserting $F_{a}(x)=2\langle\rho(z_0)\rangle^2 \partial_x^2\left[\frac{1-e^{-(a+x)}}{a+x}\right]$, one obtains
\begin{align}
    &\frac{2y^2}{\pi}\int_0^{\pi}d\varphi \sin(\varphi)[\varphi-\sin(\varphi)]\partial_x^2\left[\frac{1-e^{-x}}{x}\right]_{x=a+2y(1+\cos(\varphi))}\\
    &=-\frac{y}{\pi}\left[[\varphi-\sin(\varphi)]\partial_x\left[\frac{1-e^{-x}}{x}\right]_{x=a+2y(1+\cos(\varphi))}\right]_0^{\pi}+\frac{y}{\pi}\int_0^{\pi}d\varphi [1-\cos(\varphi)]\partial_x\left[\frac{1-e^{-x}}{x}\right]_{x=a+2y(1+\cos(\varphi))}\nonumber\\
    &=y\frac{1-(1+a)e^{-a}}{a^2}+\frac{4y}{\pi}\int_0^{\frac{\pi}{2}}d\phi \sin(\phi)^2\partial_x\left[\frac{1-e^{-x}}{x}\right]_{x=a+4y\cos(\phi)^2}\;.\nonumber
\end{align}
Gathering all the terms, one obtains
\begin{align}
 &\frac{2}{\pi}y^2 \int_0^{\pi}d\varphi \sin(\varphi)[\varphi-\sin(\varphi)]c_{a}\left(2y(1+\cos(\varphi))\right)\\
 =&y e^{-a}\left(e^{-2y}\left[I_0\left(2y\right)+I_1\left(2y\right)\right]-1\right)+y\left[1-\sqrt{\frac{a}{a+4y}}\right]-ay\left((2+a)+a\partial_a\right)\frac{1-(1+a)e^{-a}}{a^2}\nonumber\\
 &-a\left((2+a)+a\partial_a\right)\frac{4y}{\pi}\int_0^{\frac{\pi}{2}}d\phi \sin(\phi)^2\partial_x\left[\frac{1-e^{-x}}{x}\right]_{x=a+4y\cos(\phi)^2}\nonumber\\
 =&y e^{-a-2y}\left[I_0\left(2y\right)+I_1\left(2y\right)\right]-y\sqrt{\frac{a}{a+4y}}-ay\left((2+a)+a\partial_a\right)\frac{4}{\pi}\int_0^{\frac{\pi}{2}}d\phi \sin(\phi)^2\partial_x\left[\frac{1-e^{-x}}{x}\right]_{x=a+4y\cos(\phi)^2}\;.\nonumber
\end{align}
In the limit $a\to 0$, the second moment is recovered. The parametric covariance thus reads
\begin{align}
&{\rm Var}_n\left(1-\frac{v^2}{2N},{\cal D}_{z_0}(R)\right)=N^2\int_{|z_1-z|<\frac{R}{\sqrt{N}}} d^2 z_1 \int_{|z_2-z|<\frac{R}{\sqrt{N}}} d^2 z_2\,\left\langle (\rho_{0}(z_1)-\rho_v(z_1))(\rho_0(z_2)-\rho_v(z_2))\right\rangle\\
    &=2N^2\int_{|z_1-z|<\frac{R}{\sqrt{N}}} d^2 z_1 \int_{|z_2-z|<\frac{R}{\sqrt{N}}} d^2 z_2\,\left[\left\langle \rho_{0}(z_1)\rho_0(z_2)\right\rangle-\left\langle \rho_0(z_1)\rho_v(z_2)\right\rangle\right]\nonumber\\
    &=\frac{4y^2}{\pi} \int_0^{\pi}d\varphi \sin(\varphi)[\varphi-\sin(\varphi)]\left[c_{0}\left(2y(1+\cos(\varphi))\right)-c_a\left(a+2y(1+\cos(\varphi))\right)\right]\nonumber\\
    &=2y\left[\sqrt{\frac{a}{a+4y}}+(1-e^{-a})e^{-2y}\left[I_0\left(2y\right)+I_1\left(2y\right)\right]+a\left((2+a)+a\partial_a\right)\frac{4}{\pi}\int_0^{\frac{\pi}{2}}d\phi \sin^2(\phi)\partial_x\left[\frac{1-e^{-x}}{x}\right]_{x=a+4y\cos^2(\phi)}\right]\;.\nonumber
\end{align}
Finally, to simplify the expression we use the identities
\begin{align}
    \partial_x \left[\frac{1-e^{-x}}{x}\right]&=-\int_0^1e^{-\tau x}\,\tau\,d\tau\;,\\
    \frac{4}{\pi}\int_0^{\pi/2}e^{-4y\tau\cos^2{\phi}}\sin^2{\phi}\,d\phi&=
I_0(2y\tau)+I_1(2y\tau)\;,
\end{align}
which allow to obtain the final expression for the scaling form
\begin{align}
&{\rm Var}_n\left(1-\frac{v^2}{2N},{\cal D}_{z_0}(R)\right)=2y\left[\sqrt{\frac{a}{a+4y}}+{\cal V}(a,y)\right]\;,\\
&{\cal V}(a,y)=\left(1-e^{-a}\right)e^{-2y}\left[I_0(2y)+I_1(2y)\right]  -a\int_0^{1} e^{-\tau\left(2y+a\right)}\left(2+a(1-\tau)\right)\tau\left[I_0(2y\tau)+I_1(2y\tau)\right]\,d\tau\;.
\end{align}

\section{Dynamics and conjecture on universality}

In this section, we provide further details on the dynamics that are considered for our predictions on universality and the numerical simulations that are done to provide evidence for their validity. We suppose that the dynamics satisfies the following conditions:
\begin{itemize}
    \item It is global, namely all entries $X_{ij}(t)$ for $i,j=1,\cdots,N$ evolve with the same timescale.
    \item It is diffusive, i.e. $X_{ij}(t+dt)-X_{ij}(t)=R_{ij}(t)\sqrt{dt}+O(dt)$ for $i,j=1,\cdots,N$.
    \item It is stationary, i.e.
    \begin{equation}
        \forall t\in \mathbb{R}\;,\;\;P(X,t)=\left\langle\prod_{i,j=1}^N \delta\left(X_{ij}-X_{ij}(t)\right)\right\rangle=P(X,0)\equiv P(X)\;.
    \end{equation}
    \item It is isotropic, i.e.
    \begin{equation}
       \forall i,j=1,\cdots,N \;,\;\;N\frac{d}{dt}\left\langle\left|X_{ij}(t+dt)-X_{ij}(t)\right|^2\right\rangle=g_{\rm dyn}=O(1)\;,\;\;N\frac{d}{dt}\left\langle\left(X_{ij}(t+dt)-X_{ij}(t)\right)^2\right\rangle=0\;.
    \end{equation}
\end{itemize}

\subsection{Non-Hermitian Wigner ensembles}

Let us first consider non-Hermitian Wigner ensembles with i.i.d. complex entries $X_{ij}$ such that 
\begin{equation}
P(X)=\prod_{i,j=1}^N p_N(\Re(X_{ij}))p_N(\Im(X_{ij}))\;.   \label{p-iid} 
\end{equation}
In addition, we suppose that the marginal PDF $p_N(x)$ has zero average, finite second $N\left\langle|X_{ij}|^2\right\rangle=2N\int x^2 p_N(x)dx =\sigma^2$ and finite higher moments. The dynamics considered for that type of ensemble consists in randomly updating the entries with a new realization
\begin{equation}
    X_{ij}(t+dt)=\begin{cases}
     &X_{ij}(t)\;{\rm with\;probability}\;1-\gamma dt\\
     &\tilde X_{ij}\;{\rm with\;probability}\;\gamma dt
    \end{cases}\;,
\end{equation}
with $\tilde X_{ij}$ independent to $X_{ij}(t)$ with the same distribution. It is simple to check that under this dynamics, the marginal PDF of $X(t)$ is again given by Eq. \eqref{p-iid}. For this dynamics, let us analyze the expression of
\begin{equation}
 a=\lim_{N\to \infty}\pi \left\langle\sum_{n=1}^N\left|z_n\left(\frac{\tau}{N}\right)-z_n(0)\right|^2\delta^2(z_n(0)-z_0)\right\rangle\;.\label{parameters} 
\end{equation}
First, we use the identity
\begin{equation}
 dz_n={\bf l}_n^{\dagger}dX{\bf r}_n\;,
 \;\;n=1,\cdots,N\;,\label{dzdX}
\end{equation}
valid for non-Hermitian random matrices. Next, we average over the random dynamics the term
\begin{align}
    &\left\langle\left|z_n\left(\frac{\tau}{N}\right)-z_n(0)\right|^2\delta^2(z_n(0)-z_0)\right\rangle\\
    &=\left\langle\sum_{i,j=1}^N\sum_{k,l=1}^N\overline{l_{ni}}\left(X_{ij}\left(\frac{\tau}{N}\right)-X_{ij}(0)\right)r_{nj}l_{nk}\left(\overline{X_{kl}\left(\frac{\tau}{N}\right)}-\overline{X_{kl}(0)}\right)\overline{ r_{nl}}\delta^2(z_n(0)-z_0)\right\rangle\\
    &=\left\langle\sum_{i,j=1}^N|l_{ni}|^2|r_{nj}|^2\left|X_{ij}\left(\frac{\tau}{N}\right)-X_{ij}(0)\right|^2\delta^2(z_n(0)-z_0)\right\rangle\\
    &=\left\langle\sum_{i,j=1}^N|l_{ni}|^2|r_{nj}|^2\left(\left| X_{ij}\left(\frac{\tau}{N}\right)\right|^2+\left|X_{ij}(0)\right|^2-X_{ij}(0)\overline{X_{ij}}\left(\frac{\tau}{N}\right)-\overline{X_{ij}(0)}X_{ij}\left(\frac{\tau}{N}\right)\right)\delta^2(z_n(0)-z_0)\right\rangle\\
    &=\left\langle\sum_{i,j=1}^N|l_{ni}|^2|r_{nj}|^2\left(\frac{\gamma\tau}{N}\left|\tilde X_{ij}\right|^2+\left(2-\frac{\gamma\tau}{N}\right)\left|X_{ij}(0)\right|^2-2\left(1-\frac{\gamma\tau}{N}\right)\left|X_{ij}(0)\right|^2\right)\right\rangle\\
    &=\frac{2\gamma\tau}{N}\left\langle\sum_{i,j=1}^N|l_{ni}|^2|r_{nj}|^2\left|X_{ij}(0)\right|^2\delta^2(z_n(0)-z_0)\right\rangle\;.
\end{align}
In the large $N$ limit, this expression is expected to depend only on the second moment of $X_{ij}$, and summing over $n$ it converges to leading order to \cite{osman2026universality}
\begin{align}
 &\lim_{N\to\infty}\frac{2\gamma\tau}{N}\sum_{n=1}^{N}\left\langle\sum_{i,j=1}^N|l_{ni}|^2|r_{nj}|^2\left|X_{ij}(0)\right|^2\delta^2(z_n(0)-z_0)\right\rangle\\
 &=\lim_{N\to\infty}\frac{2\gamma\tau}{N}\sum_{n=1}^{N}\left\langle\sum_{i,j=1}^N|l_{ni}|^2|r_{nj}|^2\delta^2(z_n(0)-z_0)\right\rangle\left\langle\left|X_{ij}(0)\right|^2\right\rangle\\
 &=\lim_{N\to\infty}\frac{2\gamma\sigma^2\tau}{N^2}\sum_{n=1}^{N}\left\langle|{\bf l}_{n}|^2|{\bf r}_{n}|^2\delta^2(z_n-z_0)\right\rangle=2\gamma\sigma^2\tau O_1(z_0)\;,   
\end{align}
where we have used the definition of the Chalker-Mehlig mean diagonal correlator and $O_{nn}=\frac{\left(\mathbf{l}_n^{\dagger}\mathbf{l}_n\right)
\left(\mathbf{r}_n^{\dagger}\mathbf{r}_n\right)}{\left(\mathbf{l}_n^{\dagger}\mathbf{r}_n\right)
\left(\mathbf{r}_n^{\dagger}\mathbf{l}_n\right)}=|{\bf l}_{n}|^2|{\bf r}_{n}|^2$ in the normalization considered. Finally, the expression of the parameter $a$ reads for this dynamics 
\begin{equation}
 a=2\gamma\sigma^2\tau \pi O_1(z_0)=2\gamma\sigma^2\tau(1-|z_0|^2)\;.
\end{equation}
The validity of the conjecture is tested numerically by considering Bernoulli random variables
\begin{equation}
 p_N(x)=\frac{1}{2}\left[\delta\left(x-\frac{1}{\sqrt{2N}}\right)+\delta\left(x+\frac{1}{\sqrt{2N}}\right)\right]
\end{equation}
such that $\sigma^2=1$, we take $\gamma=1$, $\tau=1/2$ and $z_0=(1+i\sqrt{3})/4$. The corresponding value of $a=3/4$ while $y=R^2$. The simulations are run for a matrix size $N=400$ and for $N_{\rm it}=10^5$ iterations. The numerical data in Fig. 1 of the main text points to the fact that the matching with the scaling function is as good in this case as in the complex Ginibre ensemble (GinUE), thus providing evidence for the univerality conjecture.


\subsection{Langevin dynamics}

\begin{figure}
    \centering
    \includegraphics[width=0.8\linewidth]{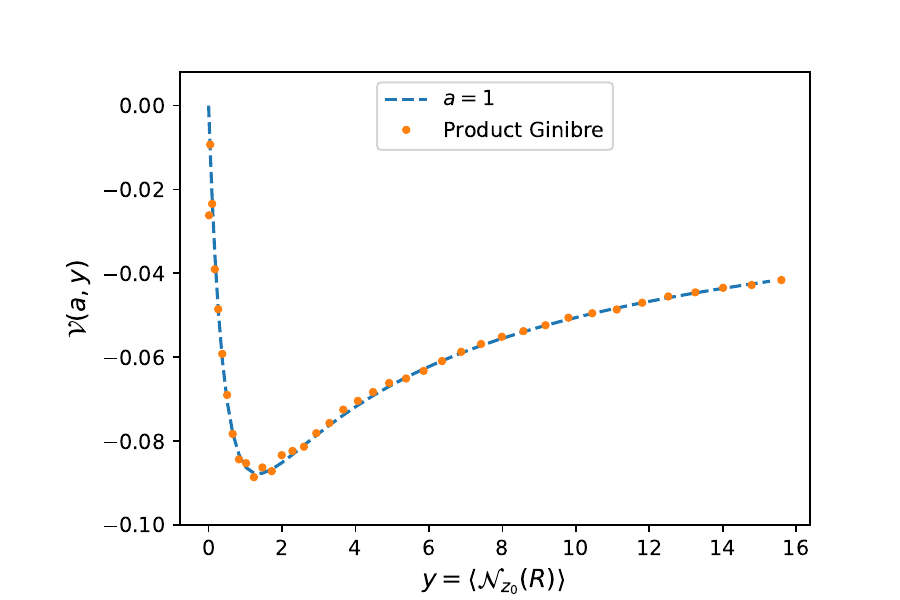}
    \caption{The rescaled parametric number covariance scaling function ${\cal V}(a,y)$ in the disk ${\cal D}_{z_0}(R)$ of centre $z_0=(1+i\sqrt{3})/4$ and varying radius $R$ is plotted as a function of the average number $y=\langle {\cal N}_{z_0}(R)\rangle$ for a Langevin evolution of the complex Ginibre ensemble. The simulation are done by perturbing a matrix $X(0)=G_1 G_2$ drawn from the product ensemble where $G_{1,2}$ are independent complex Ginibre ensemble with a matrix from the complex Ginibre ensemble (see Eq. \eqref{pert-Gin}). The good matching between the numerical data (orange dots) and the respective theoretical curve (blue dashed line) provides evidence for the validity of  the wider universality of our results. The simulations are done for matrix sizes $N=400$ and $N_{\rm it}=10^5$ realizations.}
    \label{fig:Lamgevin}
\end{figure}

Let us now consider bi-unitarily invariant ensembles of matrices such that $P(X)\propto \exp(-N{\rm Tr}W(XX^{\dagger}))$. Note that the Ginibre ensemble is the only non-Hermitian ensemble of random matrices that is both of Wigner type (i.i.d. entries) and bi-unitarily invariant with $W(u)=u$. One may easily write a Fokker-Planck equation such that the stationary solution is $P(X)$, it reads
\begin{equation}
    \partial_t P(X,t)=\frac{1}{4N}\nabla\left(2N{\rm Tr}W(XX^{\dagger})P(X,t)+\nabla P(X,t)\right)\;.
\end{equation}
The associated Langevin dynamics takes the form
\begin{equation}
    dX(t)=-\frac{1}{2}W'(X(t)X^{\dagger}(t))X(t)dt+\frac{dB(t)}{\sqrt{2N}}\;,\;\;\langle|B_{ij}(t)^2|\rangle=2t\;,\;\;\langle B_{ij}(t)^2\rangle=0\;,\label{Langdyn}
\end{equation}
where $dB(t)$ is a complex standard Brownian process on all entries of the matrix. Expanding in time for $t=\tau/N$ , one obtains
\begin{equation}
    X\left(\frac{\tau}{N}\right)-X(0)=\sqrt{\frac{\tau}{N}}G+O\left(\frac{1}{N}\right)\;,\label{pert-Gin}
\end{equation}
where $G$ is a standard complex Ginibre matrix independent from $X(0)$. Thus, at this order the perturbation is independent from the specific bi-unitary ensemble from which $X(0)$ is drawn. Using this expression and the identity in Eq. \eqref{dzdX}, one obtains 
\begin{align}
 a=\lim_{N\to \infty}\frac{\pi\tau}{N} \left\langle\sum_{n=1}^N\left|{\bf l}_n^{\dagger}G{\bf r}_n\right|^2\delta^2(z_n(0)-z_0)\right\rangle=\lim_{N\to \infty}\frac{\pi\tau}{N^2} \left\langle\sum_{n=1}^N\left|{\bf l}_n\right|^2\left|{\bf r}_n\right|^2\delta^2(z_n-z_0)\right\rangle=\tau\pi O_1(z_0)\;. 
\end{align}
Note that the left and right eigenvectors correspond to the initial random matrix $X(0)$ and not to the complex Ginibre matrix $G$ appearing in the perturbation. The average density of eigenvalues $\left\langle\rho(z_0)\right\rangle$ and the Chalker-Mehlig diagonal correlator $O_1(z_0)$ can be computed exactly in terms of the function $W(u)$ and read  \cite{belinschi2017squared}
\begin{align}
\left\langle\rho(z_0)\right\rangle&=\frac{1}{\pi}\left.\partial_u\left(u W'(u)\right)\right|_{u=|z_0|^2}=\frac{W'(|z_0|^2)+|z_0|^2 W''(|z_0|^2)}{\pi}\;,\\
O_1(z_0)&=\frac{1}{\pi |z_0|^2}\int_{|z|\leq |z_0|}\left\langle\rho(z)\right\rangle d^2z\int_{|z|\geq |z_0|}\left\langle\rho(z)\right\rangle d^2z=\frac{W'(|z_0|^2)(1-|z_0|^2 W'(|z_0|^2))}{\pi}\;.
\end{align}
Numerical simulations are done to test the validity of our conjecture in the Langevin case by considering the perturbed matrix $X(\tau/N)=X(0)+\sqrt{\tau/N}G$, where $G$ is a standard complex Ginibre matrix independent from $X(0)=G_1 G_2$, which is drawn from the product ensemble such that $G_{1,2}$ are independent standard complex Ginibre matrices. For that ensemble $W(u)=2\sqrt{u}$, such that $\langle \rho(z)\rangle=(2\pi|z|)^{-1}$ and $\langle O_1(z)\rangle=(1-|z|)/(\pi |z|)$. The simulations in Fig. \ref{fig:Lamgevin} are done for $z_0=(1+i\sqrt{3})/4$ and $\tau=1$ such that $\langle \rho(z)\rangle=\pi^{-1}$, $\langle O_1(z)\rangle=\pi^{-1}$, $y=R^2$ and $a=\tau \pi \langle O_1(z)\rangle=1$, for matrix size $N=400$ and $N_{\rm it}=10^5$ iterations. The agreement between the numerical data and the theoretical prediction provides further evidence for our conjecture.

\subsection{Unitary dynamics}

\begin{figure}
    \centering
    \includegraphics[width=0.8\linewidth]{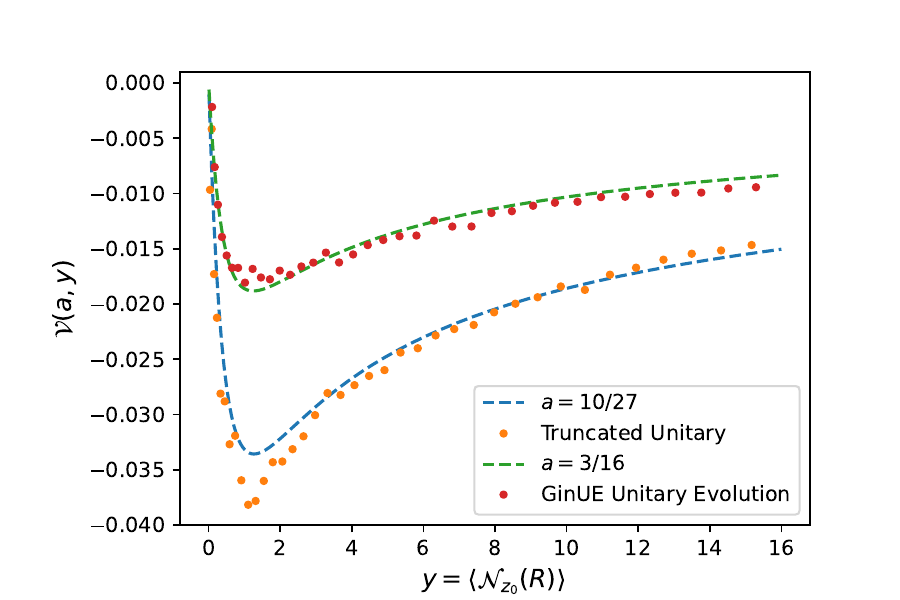}
    \caption{The rescaled parametric number covariance scaling function ${\cal V}(a,y)$ in the disk ${\cal D}_{z_0}(R)$ of centre $z_0=(1+i\sqrt{3})/4$ and varying radius $R$ is plotted as a function of the average number $y=\langle {\cal N}_{z_0}(R)\rangle$ for unitary evolutions. The simulation are done by unitary evolution of the complex Ginibre ensemble of size $N$ (red dots) and for a unitary matrix of size $M=N+L$ truncated to a size $N$ (orange dots). The good matching between the numerical data (red dots) and the respective theoretical curve (green dashed line) for unitary evolution of the Ginibre ensemble provides evidence for the validity of  the wider universality of our results. The matching for the truncated unitary ensemble between the numerical data (orange dots) and the theoretical prediction (blue dashed line) is reasonable. The discrepancy between numerical data and the theoretical prediction can be the result of finite size effects and/or spatial fluctuations: the parameter $a$ is computed at the fixed position $z_0$. The simulations are done for matrix sizes $N=400$, $M=600$ and $N_{\rm it}=10^5$ realizations.}
    \label{fig:unit}
\end{figure}

Finally, let us introduce the unitary dynamics defined (in the Itô convention) as
\begin{align}
 dX(t)=\frac{i}{\sqrt{2N}}dH(t)X(t)\;,\;\;H(t)=H^{\dagger}(t)=\frac{B(t)+B^{\dagger}(t)}{\sqrt{2}}\;,\;\;\langle|H_{ij}(t)^2|\rangle=2t\;,\;\;\langle H_{ij}(t)^2\rangle=0\;,   
\end{align}
such that $dH(t)$ a standard Brownian motion on complex Hermitian matrices and $dB(t)$ is the standard Brownian motion defined in Eq. \eqref{Langdyn}. Expanding in the short-time limit, one obtains
\begin{equation}
    X\left(\frac{\tau}{N}\right)-X(0)=i\sqrt{\frac{\tau}{N}}H X(0)+O\left(\frac{1}{N}\right)\;,
\end{equation}
where $H$ is a standard GUE matrix. Using this expression and the identity in Eq. \eqref{dzdX}, one obtains 
\begin{align}
 a=\lim_{N\to \infty}\frac{\pi\tau}{N} \left\langle\sum_{n=1}^N\left|{\bf l}_n^{\dagger}H X{\bf r}_n\right|^2\delta^2(z_n(0)-z_0)\right\rangle=\lim_{N\to \infty}\frac{\pi\tau}{N^2} \left\langle\sum_{n=1}^N\left|{\bf l}_n\right|^2 \left|{\bf r}_n\right|^2 |z_n|^2\delta^2(z_n-z_0)\right\rangle=\tau|z_0|^2\pi  O_1(z_0)\;. 
\end{align}
We perform numerical simulations for this dynamics numerically by considering $X(0)$ belonging to the complex Ginibre ensemble and $X(\tau/N)=e^{i\sqrt{\tau/N}H}X(0)$ with $H$ an independent GUE matrix. We again take value of $z_0=(1+i\sqrt{3})/4$ and $\tau=1$ such that the corresponding value of $a=\tau|z_0|^2(1-|z_0|^2)=3/16$. The good agreement between numerical data and our theoretical prediction displayed in Fig. \ref{fig:unit} provides evidence supporting our conjecture. Finally, we also apply a unitary dynamics for a unitary matrix $U_M$ of size $M=N+L$ which is then truncated it to a size $N$, i.e.
\begin{align}
    X(0)=P_N U_M P_N\;,\;\;X\left(\frac{\tau}{N}\right)=P_N e^{i\sqrt{\tau/N}H}U_M P_N\;,
\end{align}
where $H$ is a standard GUE matrix of size $M=N+L$ and $P_N$ is a projector on the first $N\times N$ block. That case corresponds to the truncated unitary ensemble for which $W(u)=-\kappa\ln(1-u)$ with $\kappa=L/N$. This yields 
\begin{align}
\left\langle\rho(z_0)\right\rangle=\frac{\kappa}{\pi\left(1-|z_0|^2\right)^2}\;,\;\;
O_1(z_0)=\frac{\kappa\left(1-(1+\kappa)|z_0|^2\right)}{\pi\left(1-|z_0|^2\right)^2}\;,\;\;a=\frac{\tau\kappa}{1+\kappa}\frac{1-(1+\kappa)|z_0|^2}{\left(1-|z_0|^2\right)^2}\;.
\end{align}
We test our conjecture on this case for $M=600$, $N=400$, $\tau=1$ and $z_0=(1+i\sqrt{3})/4$ such that $a=10/27$ with $N_{\rm it}=10^{5}$ iterations in Fig. \ref{fig:unit}, showing reasonable agreement with the prediction for the universal scaling form.

\end{document}